\documentclass[iop]{emulateapj}
\usepackage{amsmath}
\usepackage{color}
\slugcomment{Received 2009 August 21; accepted 2010 May 4}
\begin{document}
\title{Cosmic Evolution of Virial and Stellar Mass in Massive
  Early-Type Galaxies
\footnote{Based on observations with the NASA/ESA Hubble Space
  Telescope obtained at the Space Telescope Science Institute, which
  is operated by the Association of Universities for Research in
  Astronomy, Incorporated, under NASA contract NAS5-26555.}}
\shorttitle{Mass Evolution in Early-Type Galaxies}
\shortauthors{Lagattuta et al.}
\author{David J. Lagattuta\altaffilmark{1}}\email{lagattuta@physics.ucdavis.edu}\author{Christopher D. Fassnacht\altaffilmark{1}}\author{Matthew W. Auger\altaffilmark{1,2}}\author{Philip J. Marshall\altaffilmark{2}}\author{Maru{\v s}a Brada{\v c}\altaffilmark{2}}\author{Tommaso Treu\altaffilmark{2}}\author{Rapha\"{e}l Gavazzi\altaffilmark{3}}\author{Tim Schrabback\altaffilmark{4}}\author{C\'{e}cile Faure\altaffilmark{5}}\author{Timo Anguita\altaffilmark{6}}

\altaffiltext{1}{Department of Physics, University of California, Davis, 1 Shields Avenue, Davis CA 95616}
\altaffiltext{2}{Department of Physics, University of California, Santa Barbara, CA 93106}
\altaffiltext{3}{Institut d'Astrophysique de Paris, CNRS UMR 7095 \& Univ. Paris 6, 98bis Bd Arago, 75014 Paris, France}
\altaffiltext{4}{Leiden Observatory, Leiden University, Niels Bohrweg 2, 2333CA Leiden, the Netherlands}
\altaffiltext{5}{Laboratoire d'Astrophysique, \'{E}cole Polytechnique F\'{e}d\'{e}rale de Lausanne (EPFL), Observatoire de Sauverny, 1290 Versoix, Switzerland}
\altaffiltext{6}{Astronomisches Rechen-Institut, Zentrum f\"{u}r Astronomie der Universit\"{a}t Heidelberg, M\"{o}nchhofstr. 12-14, 69120 Heidelberg, Germany}

\begin{abstract}
We measure the average mass properties of a sample of 41 strong
gravitational lenses at moderate redshift (z $\sim$ 0.4 -- 0.9), and
present the lens redshift for 6 of these galaxies for the first time.
Using the techniques of strong and weak gravitational lensing on
archival data obtained from the Hubble Space Telescope, we determine
that the average mass overdensity profile of the lenses can be fit
with a power-law profile ($\Delta\Sigma \propto R^{-0.86 \pm 0.16}$)
that is within 1-$\sigma$ of an isothermal profile ($\Delta\Sigma
\propto R^{-1}$) with velocity dispersion $\sigma_v$ = 260 $\pm$ 20
km~s$^{-1}$.  Additionally, we use a two-component de Vaucouleurs+NFW
model to disentangle the total mass profile into separate luminous and
dark matter components, and determine the relative fraction of each
component. We measure the average rest frame V-band stellar
mass-to-light ratio ($\Upsilon_V = 4.0 \pm 0.6~h~
M_{\odot}/L_{\odot}$) and virial mass-to-light ratio ($\tau_V = 300
\pm 90~h~M_{\odot}/L_{\odot}$) for our sample, resulting in a
virial-to-stellar mass ratio of $M_{vir}/M_{*} = 75 \pm 25$.  Relaxing
the NFW assumption, we estimate that changing the inner slope of the
dark matter profile by $\sim$20\% yields a $\sim$30\% change in
stellar mass-to-light ratio.  Finally, we compare our results to a
previous study using low redshift lenses, to understand how galaxy
mass profiles evolve over time.  We investigate the evolution of
$M_{vir}/M_{*}(z) = \alpha(1+z)^{\beta}$, and find best fit parameters
of $\alpha = 51 \pm 36$ and $\beta = 0.9 \pm 1.8$, constraining the
growth of virial to stellar mass ratio over the last $\sim$7
Gigayears.  We note that, by using a sample of strong lenses, we are
able to constrain the growth of $M_{vir}/M_{*}(z)$ without making any
assumptions about the IMF of the stellar population.
\end{abstract}


\section{Introduction}
Observations over the last few decades have suggested that galaxies
are embedded in an extended, diffuse halo of dark matter
\citep[e.g.][]{pet78,whi78,rub79,blu84}, which contributes up to 95
percent of the total mass \citep{hoe05,jia07,avi08}.  Because of the
overwhelming fraction of dark matter, numerical simulations dealing
with the formation and assembly of galaxy mass often focus on dark
matter alone \citep[e.g., the Via Lactea Simulation;][]{kuh08}, and
these simulations have made very specific predictions about the
overall profile of galaxy-sized mass distributions
\citep[e.g.,][]{nav97,moo98,jin00,sto06,die07,sch08}.  However, when
compared to observations, these simulations often fail to accurately
recreate the observed shape of galaxy mass profiles
\citep{sal01,gen04,sim05,gen07,kuz08,sei08,san08}. This disagreement
is attributed largely to the complex physics of baryon interaction
(e.g. frictional dissipation and radiative cooling) which is still not
well understood and is difficult to model.

There are many possible methods capable of measuring the distribution
of matter on small ($<$ 1 Mpc) scales, and techniques used for
characterizing a radial mass profile include measuring the rotation
curves from planetary nebulae \citep{rom03,arn04,mer06,del08,nap09},
kinematics of stellar populations \citep{ber94,cap06,van08} and
\ion{H}{1} gas \citep{uso03,jac04,and06,mat08}, or the temperature of
X-ray emitting gas \citep{hum06,chu08}.

In this paper, we use a powerful alternative technique: gravitational
lensing \citep[e.g.,][]{bra96,fis00,wil01,kle06,par07}.  The power of
lensing is due to the fact that the technique is able to directly
trace the total mass (luminous + non-luminous) enclosed within a given
radius without needing to make any assumptions about the dynamical
state of the mass in question.  In addition, lensing does not rely on
the presence of kinematic tracers, which are often only visible in
very low-redshift galaxies, and even then are typically not present in
the outer halo regions where dark matter dominates.

Furthermore, since we analyze a sample of strong lenses, we are able
to combine the mass constraints determined from both strong lensing
and weak lensing.  This allows us to probe the total mass distribution
over a much wider range of physical distances than by using weak
lensing alone, and also allows us to constrain the properties of
stellar mass distribution without selecting a specific stellar initial
mass function (IMF).  Instead, we rely on directly-derived properties:
lensing-inferred mass and total luminosity (although we do make an
assumption about the shape of the dark matter halo).

For early-type galaxies, several lensing-based studies have shown the
typical shape of the mass density profile to be consistent with an
isothermal ($\rho(r) \propto r^{-2}$) model, given the uncertainties
on the measurements, between distances of $\sim 50-300 ~h^{-1}$ kpc
\citep{she04,man06}.  However, due to the low signal-to-noise ratio
(SNR) on the lensing signal obtained from an individual galaxy, these
studies rely on stacking a large number of galaxies into a single
sample to determine the average profile of the entire stack.  For this
work, we utilize deep imaging from the Hubble Space Telescope (HST).
In general, space-based data will have a much higher density of
background galaxies ($n_s$) when compared to data taken from the
ground over an identical exposure time.  This increased background
density count not only enables us to detect a weak lensing signal much
closer to the center of the lensing galaxy, but also allows for a
significant detection with fewer stacked galaxies ($\rm N_{lens}$),
because the expected weak lensing SNR scales as $\sqrt{{\rm
    N_{lens}}~n_s}$.  The ability to work on a small sample size is
particularly useful, given the relative paucity of known strong lenses
at all redshifts.

In this work, we are able to extend the results of \citet[hereafter
  G07]{gav07}, who conducted a similar mass profile study on a small
(22 lens) sample of low redshift (z $\sim$ 0.2) early-type strong
gravitational lenses collected from the Sloan Lens ACS Survey
\citep[SLACS; see e.g.][]{bol06,bol08} over a wide range of physical
distances ($\sim 3 - 300 ~h^{-1}$ kpc).  From this work, it was shown
that, like their non lensing counterparts, the total mass density
profile of strong lensing ellipticals could be described by a roughly
isothermal model (although they note that the comparison between the
data and the model was done without a formal fit).  This was thought
to be due to the combination of baryonic and dark matter, the
so-called Bulge-Halo Conspiracy, in which a relatively steep luminous
matter profile and a shallower dark matter profile combine in such a
way that the total mass distribution of the galaxy can be well
approximated by an isothermal model \citep{tre06}.  By comparing this
data set to our moderate redshift sample, we thus explore the
evolutionary trends of massive early-type galaxies.
 
This paper is organized as follows: In Section 2, we describe our lens
sample, the lens selection criteria, and the data reduction pipeline
used for this work.  At this point, the interested reader may turn to
the appendix, where we describe our techniques for analyzing weak
lensing shear, paying specific attention to signal to noise
optimization and the removal of systematic errors.  In Section 3, we
use a two-component model to determine the contributions of the
stellar and dark matter components to the total profile.  We present
all of our results in Section 4, the reader interested in the
science may skip directly to this section.  We discuss the results and
compare them to the SLACS results in Section 5, looking for
evolutionary differences between the samples.  Finally, we summarize
our results in Section 6.

Throughout this work, we assume a flat cosmology with $H_0 = 100 ~h$
km s$^{-1}$ Mpc$^{-1}$, $\Omega_M = 0.3$, and $\Omega_\Lambda = 0.7$.
All magnitudes presented in this paper are AB magnitudes.

\section{Data}
\subsection{Imaging Data}
For this analysis, we focus on a sample of 41 strong gravitational
lenses at moderate redshift ($z \sim 0.4 - 0.9, ~z_{\rm median} \sim
0.6$).  A majority of the sample was observed as part of the CASTLES
program\footnote{http://cfa-www.harvard.edu/castles/}, but it also
contains lenses found in the COSMOS
survey\footnote{http://cosmos.astro.caltech.edu/} \citep{fau08} and
the Extended Groth Strip \citep{mou07}, as well as targeted exposures
of individual lens systems.  Unlike other data sets that have been
selected in some uniform way (such as the SLACS lenses), our lens
sample is drawn from a variety of sources and selected simply for
being lenses with early-type morphologies in the desired redshift
range.  However, we do note that systems known to be strongly affected
by the presence of a galaxy cluster (Q0957+561, SDSS1004+4112, and
B2108+213) were excluded from the sample.  A full list of the lenses
can be seen in Table~\ref{tbl:lens}.

\begin{deluxetable*}{lrcllrrrr}
\tabletypesize{\scriptsize}
\tablecaption{Lens System Data. \label{tbl:lens}}
\tablehead{
\colhead{Lens Name} & \colhead{Program} & \colhead{Exp. Time} & \colhead{$z_l$\tablenotemark{a,b}} & \colhead{$z_s$\tablenotemark{b}} & \colhead{$D_l$} & \colhead{$\overline{w}(z_l)$\tablenotemark{c}} & \colhead{$\Sigma_{\rm crit}$} & \colhead{References}\\ \colhead{} & \colhead{ID} & \colhead{($s$)} & \colhead{} & \colhead{} & \colhead{($h^{-1}$ Mpc)} & \colhead{} & \colhead{($M_\odot$ pc$^{-2}$)} & \colhead{}
}
\startdata
B0218+357       &9450   &4320.0  &0.68     &0.96        &1020   &0.320  &5097  &[1],[2],[3]	\\
SDSS0246-0825	&9744	&2288.0  &0.724	   &1.68	&1045	&0.293	&5035  &[4]	\\	
CFRS03P1077	&9744	&2296.0	 &0.938	   &2.941	&1137	&0.188	&8965  &[5]	\\	
HE0435-1223	&9744	&1445.0	 &0.454	   &1.689	&842	&0.486	&3702  &[6],[7]	\\	
B0445+128	&9744	&5228.0	 &0.557	   &\nodata	&931	&0.406	&3969  &[8]	\\	
B0631+519	&9744	&2446.0	 &0.62	   &\nodata	&979	&0.360	&4267  &[9]	\\	
J0816+5003	&9744	&2440.0	 &\nodata  &\nodata	&825	&0.500	&3678  &[10]	\\	
B0850+054	&9744	&2296.0	 &0.59	   &3.93	&957	&0.381	&4112  &[11]	\\	
SDSS0903+5028	&9744	&2444.0	 &0.388	   &3.605	&761	&0.552	&3645  &[12]	\\	
SDSS0924+0219	&9744	&2296.0	 &0.394	   &1.524	&763	&0.550	&3644  &[13],[14]	\\	
J1004+1229	&9744	&2296.0	 &0.95	   &2.65	&1141	&0.183	&9397  &[15]	\\	
HE1113-0641	&9744	&1062.0	 &0.75	   &1.235	&1059	&0.278	&5298  &[16]	\\	
Q1131-1231	&9744	&1980.0	 &0.295	   &0.658	&635	&0.646	&3803  &[17]	\\	
SDSS1138+0314	&9744	&2296.0	 &0.45	   &2.44	&831	&0.495	&3685  &[18],[19]	\\	
SDSS1155+6346	&9744	&1748.0	 &0.176	   &2.89	&430	&0.780	&4782  &[20]	\\	
SDSS1226-0006	&9744	&2296.0	 &0.52	   &1.12	&899	&0.435	&3841  &[18],[19]	\\	
B1608+656       &10158  &9744.0  &0.63     &1.39        &987    &0.353  &4774  &[21],[22]	\\ 
WFI2033-4723	&9744	&2085.0	 &0.66	   &1.66	&1007	&0.333	&4516  &[18],[23]	\\	
COSMOS5857+5949	&9822	&2028.0	 &0.39	   &\nodata	&763	&0.550	&3646  &[24]	\\
COSMOS5914+1219	&9822	&2028.0	 &1.13	   &\nodata	&1169	&0.148	&15755 &[24]	\\
COSMOS5921+0638	&9822	&2028.0	 &0.551	   &3.15	&926	&0.411	&3948  &[24],[25]	\\
COSMOS5941+3628	&9822	&2028.0	 &0.88	   &\nodata	&1124	&0.204	&7834  &[24]	\\
COSMOS5947+4752	&10092	&2028.0	 &0.345	   &\nodata	&706	&0.595	&3681  &[24],[26]	\\
COSMOS0012+2015	&9822	&2028.0	 &0.378	   &\nodata	&749	&0.562	&3649  &[24],[26]	\\
COSMOS0013+2249	&9822	&2028.0	 &0.346	   &\nodata	&707	&0.594	&3680  &[24],[26]	\\
COSMOS0018+3845	&9822	&2028.0	 &0.71	   &\nodata	&1037	&0.301	&4908  &[24]	\\
COSMOS0038+4133	&10092	&2028.0	 &0.738	   &\nodata	&1121	&0.208	&7583  &[24],{\bf [29]}	\\	
COSMOS0047+5023	&9822	&2028.0	 &0.87	   &\nodata	&1105	&0.226	&6729  &[24]	\\
COSMOS0049+5128	&9822	&2028.0	 &0.337	   &\nodata	&695	&0.603	&3694  &[24],[26]	\\	
COSMOS0050+4901	&10092	&2028.0	 &0.960	   &\nodata	&1144	&0.179	&9795  &[24],[26]	\\	
COSMOS0056+1226	&9822	&2028.0	 &0.361	   &0.81	&727	&0.579	&3661  &[24],[26],{\bf [29]}	\\
COSMOS0124+5121	&9822	&2028.0	 &0.84	   &\nodata	&1101	&0.231	&6546  &[24]	\\
COSMOS0211+1139	&9822	&2028.0	 &0.920	   &\nodata	&1124	&0.204	&7834  &[24],{\bf [29]}	\\
COSMOS0216+2955	&9822	&2028.0	 &0.608	   &\nodata	&1013	&0.326	&4588  &[24],{\bf [29]}	\\
COSMOS0227+0451	&10092	&2028.0	 &0.89	   &\nodata	&1121	&0.208	&7583  &[24]	\\
COSMOS0254+1430	&10092	&2028.0	 &0.417    &0.779	&825	&0.500	&3678  &[24],{\bf [29]}	\\
J095930.93+023427.7 &9822   &2028.0  &0.892    &\nodata     &1122   &0.207  &7150  &[27],{\bf [29]}    \\
J100140.12+020040.9 &9822   &2028.0  &0.879    &\nodata     &1117   &0.213  &6987  &[27],{\bf [29]}    \\
``Anchor''  	&10134	&2100.0	 &0.463	   &\nodata	&845	&0.483	&3707  &[28]	\\
``Cross''   	&10134	&2100.0	 &0.810	   &3.40	&1088	&0.246	&6060  &[28]	\\
``Dewdrop'' 	&10134	&2100.0	 &0.580	   &0.982	&950	&0.389	&4068  &[28]	\\
\enddata

\tablenotetext{a}{$\rm Lenses$ with no known $z_l$ were placed at a fiducial redshift z = 0.6.}
\tablenotetext{b}{For the COSMOS lenses, redshift values with 3 significant figures were determined spectroscopically, and redshift values with 2 significant figures are determined photometrically, according to the catalog of \citet{ill09}.}
\tablenotetext{c}{$\overline{w}(z_l)$ is the weight value associated with each lens field, described in the appendix.}
\tablenotetext{~}{References: {\bf[1]}: \citet{pat93}, {\bf[2]}: \citet{bro93}, {\bf[3]}: \citet{law96}, {\bf[4]}: \citet{ina05}, {\bf[5]}: \citet{cra02}, {\bf[6]}: \citet{wis02}, {\bf[7]}: \citet{mor05}, {\bf[8]}: \citet{arg03}, {\bf[9]}: \citet{yor05}, {\bf[10]}: \citet{leh01}, {\bf[11]}: \citet{big03}, {\bf[12]}: \citet{joh03}, {\bf[13]}: \citet{ina03}, {\bf[14]}: \citet{eig06a}, {\bf[15]}: \citet{lac02}, {\bf[16]}: \citet{bla08}, {\bf[17]}: \citet{slu03}, {\bf[18]}: \citet{eig06b}, {\bf[19]}: \citet{ina08}, {\bf[20]}: \citet{pin04}, {\bf[21]}: \citet{mye95}, {\bf[22]}: \citet{fas96}, {\bf[23]}: \citet{mor04}, {\bf[24]}: \citet{fau08}, {\bf[25]}: \citet{ang09}, {\bf[26]}: C. Faure et al. in preparation, {\bf[27]}: \citet{jac08}, {\bf[28]}: \citet{mou07}, {\bf[29]}: This paper}
\end{deluxetable*}

Each system in the sample was chosen based on the existence of
publicly available HST images, subject to the following criteria: the
system needed to be observed using the Wide Field Channel (WFC) of the
Advanced Camera for Surveys \citep[ACS;][]{for98}, and the total
exposure needed to be at least $\sim$ 1800 seconds through the F814W
filter.  This exposure time yields background densities of $\sim$ 70
galaxies ${\rm arcmin}^{-2}$ or more.  

The imaging data for all the lenses were obtained through the
Multimission Archive at STScI
(MAST\footnote{http://archive.stsci.edu/}), from programs G0-9744
(CASTLES: PI Kochanek), GO-9822 and 10092 (COSMOS: PI Scoville),
GO-10134 (EGS: PI Davis), GO-10158 (B1608+656: PI Fassnacht), and
GO-9450 (B0218+357: PI Jackson).  The lens systems that had been
specifically targeted by HST (B1608+656, B0218+357, and the CASTLES
lenses) are located at the WF1 target point, whereas the
serendipitously discovered COSMOS and EGS lenses appear at random
positions in the ACS field of view.

In addition to the F814W data, some lens systems have secondary
imaging in either the F555W or F606W filter.  These extra data could
prove to be useful in future refinements to this study, especially in
using color selection to reject foreground interlopers from the
population of background sources (see the appendix).  However, we do
not use the information in our present weak lensing analysis: a
significant fraction of our sample has only F814W imaging, and for
systems that do have multi-band data, the redshift/color relationship
can be highly degenerate when dealing with only two filters.  Thus, we
do not expect significant improvement in the signal.

The raw data were processed through a reduction pipeline created for
the HST Archive Galaxy-scale Gravitational Lens Survey (HAGGLeS,
P.J. Marshall et al., in preperation), which we briefly describe.
First, each individual raw exposure of a given lens system is
calibrated, using the \textit{calacs} package in
STSDAS\footnote{STSDAS is a product of the Space Telescope Science
  Institute, which is operated by AURA for NASA}, a software system
built on top of IRAF\footnote{IRAF (Image Reduction and Analysis
  Facility) is distributed by the National Optical Astronomy
  Observatory, which is operated by the Association of Universities
  for Research in Astronomy, Inc., under cooperative agreement with
  the National Science Foundation.}.  Once calibrated, the exposures
are aligned and combined into a single stacked image using
multidrizzle \citep{koe02}.  Since even slight misalignments can
introduce systematic errors in the weak lensing signal, the HAGGLeS
pipeline refines the image alignment derived from the astrometric
header by cross-correlating the positions of bright, well-defined
objects (bright unsaturated stars, nebular knots in spiral galaxies,
etc.) in each exposure in order to look for residual shift or rotation
misalignments.  These residual shifts are then fed to multidrizzle,
and this process is repeated until the shift refinements become
negligibly small.  After final alignment and combination, the
composite output image is resampled in multidrizzle from the native
scale of 0.05\arcsec pixel$^{-1}$ to one of 0.03\arcsec pixel$^{-1}$,
using the ``Square'' interpolation kernel.  This is done primarily to
better sample the ACS instrumental point-spread function (PSF).
Lastly, the resampled image is registered to a common World Coordinate
System using positions from the USNO-B1 catalog \citep{mon03}.

After processing the images, we create initial photometric galaxy
catalogs for each field using SExtractor \citep{ber96}.  As a goal of
this study is to compare our results as closely as possible to those
in G07, we use their SExtractor parameter set for our analysis.  The
parameters can be seen in Table \ref{tbl:sex}.  These parameters are
selected to optimize detection of a suitably high number of small, dim
objects, while also deblending close neighbors.  This leads to a large
number of spurious detections but the ``false positives'' are rejected
during the analysis by applying a series of cuts to the data.  This
procedure is discussed more fully in the appendix.

%---------------------------------------------------------------------
\begin{deluxetable}{lc}
\tablecaption{Relevant SExtractor Parameters. \label{tbl:sex}}
\tablehead{
\colhead{Parameter Name} & \colhead{Value}
}
\startdata
\verb"DETECT_MINAREA"	& 10	   \\
\verb"DETECT_THRESH"	& 1.1 	   \\
\verb"DEBLEND_NTHRESH"	& 64	   \\
\verb"DEBLEND_MINCONT"	& 0.005	   \\
\verb"PHOT_AUTOPARAMS"  & 2.5, 3.5 \\
\verb"PHOT_FLUXFRAC"    & 0.5      \\
\verb"BACK_SIZE"	& 64	   \\
\verb"BACK_FILTERSIZE"	& 3	   \\
\verb"BACKPHOTO_TYPE"	& GLOBAL   \\
\enddata
\end{deluxetable}
%---------------------------------------------------------------------

\subsection{Spectroscopic Data}
While the determination of the weak lensing signal uses a generalized
description of the redshift distribution of the background objects,
the additional inclusion of the strong lensing information in the
analysis (\S3.3) is only possible if the redshifts of both the lens
and the multiply-imaged background have been measured.  Thus, to
improve the utility of the lens sample, we obtained spectroscopic data
on some of the best lens candidates from the COSMOS sample using the
Low-Resolution Imaging Spectrograph \citep[LRIS;][]{lris} on the Keck
I Telescope.  The observations were conducted on UT 2009 Feb 22 and 23
in good conditions, with seeing ranging from 0\farcs8 to 1\arcsec.
Spectra were obtained for the J095930.93+023427.7, 0038+4133,
0050+4901, 0056+1226, 0211+1139, J100140.12+020040.9, 0216+2955, and
0254+1430 lens systems, where 5930+3427, 0056+1226,
J100140.12+020040.9, 0211+1139, and 0216+2955 were observed through
slitmasks and the remaining systems were observed using a
1\arcsec\ longslit.  The exposure times were set by the F814W
magnitude of the lensing galaxy and ranged from 1200 to 7200~s.  The
lensed source galaxies were much fainter than the lenses and, thus,
the chosen exposure times were not sufficient to measure redshifts
unless strong emission features or breaks in the spectrum were
observed.  In cases where we could not determine a redshift for the
source galaxy, strong lensing constraints could not be obtained.
However, these lenses are still useful for, and are included in, the
weak lensing analysis.  The data were reduced using custom Python
scripts that performed the flat-field corrections, wavelength
calibrations, rectifications, and extractions of the spectra.  The
extracted spectra were examined for multiple emission and/or
absorption features.  We were able to determine lens redshifts for six
of the seven systems that were targeted (the slit for 0056+1226 was
placed only over the lensed source because the lens redshift was
previously known), but only two source redshifts.  Thus, the new
redshifts for COSMOS systems resulting from this work are:
J095930.93+023427.7 ($z_{\rm lens} = 0.892$), 0038+4133 ($z_{\rm lens}
= 0.738$), 0056+1226 ($z_{\rm src} = 0.808$, based mostly on the
4000\AA\ break), J100140.12+020040.9 ($z_{\rm lens} = 0.879$),
0211+1139 ($z_{\rm lens} = 0.92$, based mostly on the 4000\AA\ break),
0216+2955 ($z_{\rm lens} = 0.608$), and 0254+1430 ($z_{\rm lens} =
0.417$, $z_{\rm src} = 0.779$).

\section{Disentangling Mass and Light}
After reducing the data, we perform a full weak lensing analysis on
the galaxy fields to infer the average mass properties of our lenses.
The interested reader can find the full details of this analysis in
the appendix.  To summarize however, we measure the shapes of all
background galaxies, apply a series of data cuts to remove
contaminants, and employ a system of tests to minimize systematic
uncertainty.  We then convert these galaxy shape measurements into a
measure of mass overdensity as a function of radius
($\Delta\Sigma(R)$), which we can compare with specific models to
infer other properties of the lenses.
The $\Delta\Sigma(R)$ profile reflects the \textit{total} mass
distribution of the lens galaxy.  We can investigate the \textit{dark
  matter} contribution by jointly modeling the luminous and dark
components, constrained by the lens galaxy surface brightness
profiles.  To disentangle the luminous stellar mass profile from its
surrounding dark matter halo, we employ a simple two-component mass
model, whereby the stellar mass profile traces the light profile, and
the dark profile exists as a single dark matter halo, centered on the
galaxy itself (a one-halo central term in the framework of the ``halo
model'').  This simplifying assumption explicitly excludes any mass
contribution from an underlying group/cluster halo, or nearby
satellite galaxies (the two-halo term), but we feel it is a well
warranted assumption, because only a small fraction of massive
elliptical galaxies ($\la$ 15\%) are located away from the center of
their host halos \citep{man06}.

\subsection{Luminous Component}
\begin{figure}
\plotone{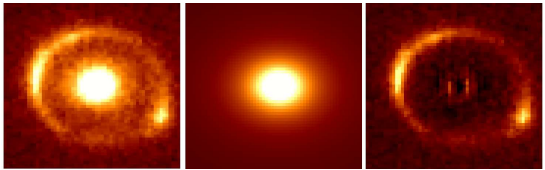}
\caption{Example of our stellar light profile model using GALFIT.
  Left: original data (COSMOS 0038+4133), showing the lensing galaxy
  and lensed Einstein ring.  Middle: model of the lens galaxy.  Right:
  residual image.  Aside from a slight undersubtraction in the core,
  the lens galaxy is well fit by the model.}
\label{fig:galfit}
\end{figure}

We begin our process of modeling stellar light by using
GALFIT\footnote{http://users.ociw.edu/peng/work/galfit/galfit.html}
\citep{pen02} to fit a de Vaucouleurs profile \citep{dev48} to the
F814W images of each strong lensing galaxy, carefully masking out any
contaminating light from nearby satellites or (more importantly) the
strongly lensed images.  In most cases, this masking is achieved by
running GALFIT on a small cutout of the galaxy (typically a square 2.5
times the galaxy's \verb"FWHM" parameter {--} as determined by
SExtractor {--} on a side).  Nearly all of the light from our
moderate-redshift lenses is contained well within the lensing Einstein
radius, making it possible to exclude all of the lensed background
structure without excluding significant portions of the light profile
of the foreground galaxy.  In the few cases where this cutout method
is not feasible (e.g. compact 4-image quasar lenses such as
SDSS1138+0314 or HE1113-0641) we generate custom GALFIT masks to block
out the remaining contaminating light.  An example of the GALFIT
modeling can be seen in Figure \ref{fig:galfit}.

%-----------------------------------------------------------------------
%-----------------------------------------------------------------------

By explicitly fitting with a de Vaucouleurs profile, we obtain two
parameters for each galaxy that can be used in the mass profile
modeling: an effective radius ($R_{\mathrm{e}}$) and an F814W (I-band)
magnitude ($m_I$).  We first convert the I-band apparent magnitudes
into F555W (V-band) absolute magnitudes ($M_V$) by calculating
K-corrections using the elliptical galaxy template from the
Kinney-Calzetti spectral atlas \citep{kin96}, and correcting for
galactic extinction using the \citet{sch98} dust maps.  This is done
so that we may more readily compare our results to both G07 and
previous mass-to-light ($M/L$) studies.  To homogenize the sample, we
passively evolve all V-band luminosities to a fiducial redshift $z =
0.6$, using the relation \citep{gri09}:
\begin{equation}
\frac{d \log L_{V}}{dz} \approx 0.6 \pm 0.05
\end{equation}

We also convert the total luminosity into effective surface
brightness ($I_e$) using the equation:
\begin{equation}
I_e = \frac{L_V~\kappa^{2n}}{2 \pi R_{\mathrm{e}}^2~n~e^\kappa~\Gamma(2n)}  
\label{eqn:mag2Ie}
\end{equation}
where $n$ and $\kappa$ are the general S\'{e}rsic parameters (for the
de Vaucouleurs profile, $n = 4$, $\kappa = 7.67$).  %, and $\Gamma()$
The relevant parameters obtained through the photometric modeling are
in Table \ref{tbl:photparams}.

We model the stellar contribution to the mass profile as:
\begin{equation}
\Sigma_* = \Upsilon_VI_e\exp{\left[-7.67\left(\left(\frac{R}{R_e}\right)^{1/4}~-~1 \right)\right]}
\label{eqn:starmass}
\end{equation}
where $\Sigma_*$ is the stellar mass surface density, and $\Upsilon_V$
is the rest-frame V-band stellar mass to light ratio ($M_*/L_V$).
Since both $I_e$ and $R_e$ are fixed by the GALFIT model for any given
galaxy, this stellar mass profile has the benefit of being described
by only a single free parameter, $\Upsilon_V$.

%-----------------------------------------------------------------------
\begin{deluxetable*}{lrrrrrcrrr}
\tabletypesize{\scriptsize}
\tablecaption{Photometric Lens Data. \label{tbl:photparams}}
\tablehead{
\colhead{Lens Name} &\colhead{$R_{Ein}$} &\colhead{$R_e$} &\colhead{m$_I$} &\colhead{K-corr} &\colhead{E(B-V)} &\colhead{M$_V$ - 2.5log$_{10}$($h$)}  &\colhead{I$_e$} &\colhead{q} &\colhead{PA}  \\ \colhead{} & \colhead{($\arcsec$)} & \colhead{($\arcsec$)} & \colhead{} & \colhead{} & \colhead{} &\colhead{} & \colhead{($L_{\odot}$ kpc$^{-2}$)} & \colhead{} & \colhead{}
}

\startdata
B0218+357       &0.17     &0.37   &19.89  &-0.19  &0.0680  &-20.10  & 123.62  &0.97  & 66.02\\
SDSS0246-0825	&0.6      &0.99   &19.81  &-0.10  &0.0266  &-21.78  & 557.53  &0.53  & 24.88\\
CFRS03P1077	&1.05     &1.89   &20.30  & 0.42  &0.0983  &-23.46  &  54.24  &0.66  & 32.40\\
HE0435-1223     &1.21     &1.12   &19.17  &-0.49  &0.0590  &-21.65  &  99.52  &0.79  &-83.16\\
B0445+128	&0.68     &2.29   &20.09  &-0.38  &0.3837  &-21.35  &  12.94  &0.71  &-54.21\\
B0631+519	&0.58     &0.84   &20.30  &-0.29  &0.0890  &-21.65  & 104.31  &0.81  &  8.35\\
J0816+5003	&2.5      &0.94   &18.54  &-0.51  &0.0466  &-22.94  & 293.74  &0.86  & 46.23\\
B0850+054	&0.34     &0.25   &21.73  &-0.34  &0.0615  &-20.60  & 478.79  &0.42  & 55.28\\
SDSS0903+5028	&1.5      &0.55   &19.34  &-0.58  &0.0245  &-20.93  & 293.68  &0.93  &-69.39\\
SDSS0924+0219	&0.88     &0.58   &19.38  &-0.58  &0.0551  &-20.90  & 248.06  &0.91  & 52.62\\
J1004+1229	&0.77     &0.51   &21.80  & 0.45  &0.0372  &-21.88  & 165.29  &0.38  & 64.87\\
HE1113-0641	&0.44     &2.35   &18.16  &-0.05  &0.0386  &-21.72  & 168.03  &0.76  &-59.19\\
Q1131-1231	&1.9      &3.63   &16.78  &-0.68  &0.0352  &-22.71  &  55.83  &0.78  & 10.93\\
SDSS1138+0314	&0.67     &0.42   &20.12  &-0.50  &0.0196  &-20.60  & 286.33  &0.79  &  7.79\\
SDSS1155+6346	&0.98     &0.52   &17.60  &-0.80  &0.0141  &-20.48  & 898.45  &0.59  & 36.23\\
SDSS1226-0006	&0.63     &0.46   &18.98  &-0.42  &0.0233  &-21.32  & 230.87  &0.28  &-34.38\\
B1608+656       &1.14     &0.89   &19.51  &-0.28  &0.0310  &-20.24  &26.09    &0.44  &-13.98\\
WFI2033-4723	&1.17     &0.73   &20.25  &-0.22  &0.0461  &-21.78  & 140.47  &0.82  & 41.79\\
COSMOS5857+5949	&2.15     &1.23   &19.63  &-0.58  &0.0193  &-20.67  &  45.30  &0.60  & 58.41\\
COSMOS5914+1219 &1.86     &1.27   &22.11  & 0.75  &0.0202  &-22.33  &  29.20  &0.69  &-22.72\\
COSMOS5921+0638 &0.8      &0.33   &20.62  &-0.39  &0.0205  &-20.77  & 366.25  &0.84  &-62.64\\
COSMOS5941+3628 &1.21     &0.83   &20.90  & 0.32  &0.0191  &-22.43  & 120.08  &0.90  & 69.28\\
COSMOS5947+4752 &2.55     &0.35   &19.96  &-0.64  &0.0209  &-19.97  & 365.42  &0.89  & 89.20\\
COSMOS0012+2015 &0.9      &0.51   &19.42  &-0.59  &0.0185  &-20.84  & 321.34  &0.64  & 68.23\\
COSMOS0013+2249 &1.65     &2.02   &18.33  &-0.63  &0.0178  &-21.69  &  52.91  &0.82  &-54.76\\
COSMOS0018+3845 &0.4      &0.32   &23.13  &-0.13  &0.0187  &-19.20  &  61.48  &0.68  & 55.42\\
COSMOS0038+4133 &0.74     &1.11   &20.36  & 0.29  &0.0186  &-22.31  &  82.28  &0.74  & 88.38\\         
COSMOS0047+5023	&0.7      &1.33   &20.16  & 0.19  &0.0182  &-23.06  &  84.67  &0.80  &-56.86\\
COSMOS0049+5128	&2.22     &0.31   &20.09  &-0.64  &0.0182  &-19.81  & 428.99  &0.75  & 22.46\\
COSMOS0050+4901	&1.9      &0.74   &21.21  & 0.48  &0.0189  &-22.53  & 140.32  &0.73  &-65.24\\
COSMOS0056+1226	&1.2      &0.76   &18.92  &-0.62  &0.0164  &-21.16  & 210.72  &0.90  &-61.32\\
COSMOS0124+5121	&0.84     &0.27   &22.33  & 0.17  &0.0188  &-20.79  & 271.75  &0.65  & 38.48\\
COSMOS0211+1139	&3.2      &1.66   &20.50  & 0.32  &0.0168  &-23.05  &  48.23  &0.61  &-83.30\\
COSMOS0216+2955	&1.96     &0.99   &19.98  &-0.20  &0.0185  &-21.70  &  84.55  &0.81  & 62.14\\
COSMOS0227+0451	&1.62     &0.71   &21.37  & 0.29  &0.0176  &-22.05  & 112.95  &0.60  &-10.32\\
COSMOS0245+1430	&1.54     &1.70   &18.63  &-0.51  &0.0191  &-22.26  &  72.62  &0.59  & 31.75\\
J095930.93+023427.7 &0.89     &0.95   &21.26  & 0.28  &0.0191  &-22.07  &  63.99  &0.69  & -2.33\\
J100140.12+020040.9 &0.79     &0.29   &21.71  & 0.26  &0.0180  &-21.58  & 446.53  &0.83  &-71.80\\
``Anchor''     	&1.1      &0.49   &19.88  &-0.49  &0.0103  &-20.93  & 285.63  &0.92  &-56.16\\
``Cross''       &1.22     &0.90   &20.29  & 0.09  &0.0085  &-22.60  & 137.47  &0.79  & 78.92\\
``Dewdrop''     &0.76     &0.50   &19.96  &-0.35  &0.0089  &-21.59  & 314.46  &0.85  & 79.14\\
\enddata

\tablenotetext{~}{Galaxy parameters obtained through GALFIT photometry
  fitting.  $R_{Ein}$ is the Einstein radius of the lens (determined
  outside of GALFIT), and $R_e$ is the effective (half-light) radius
  of the de Vaucouleurs profile. m$_I$ is the modeled F814W apparent
  magnitude of the galaxy, K-corr represents the F814W to F555W
  K-correction, E(B-V) is the galactic extinction correction term,
  M$_V$ is the absolute F555W magnitude (scaled by the Hubble
  parameter), and I$_e$ is the effective surface brightness of the
  galaxy's de Vaucouleurs profile, as determined from M$_V$.  Finally,
  q represents the axis ratio of the galaxy, and PA is its position
  angle, defined to be north through east.  While these final two
  parameters are reported as part of the GALFIT modeling, they are not
  used in the mass modeling of \S3.}
\end{deluxetable*}
%-----------------------------------------------------------------------

\subsection{Dark Component}
For the dark matter, we assume a functional form of the Navarro, Frenk
and White \citep[NFW;][]{nav97} profile.  The projected, two
dimensional NFW mass density profile ($\Sigma_{\rm{NFW}}$) can be
described generally by:
\begin{equation}
\Sigma_{NFW} \propto R_s\delta_c\rho_c 
\label{eqn:darkmass}
\end{equation}
\citep[e.g.,][]{wri00}, where $R_s$ is the NFW ``scale radius'',
$\rho_c$ is the critical density of the universe, and $\delta_c$ =
($\Delta/3$)$c^3$/[ln($1 + c$) - $\frac{c}{1+c}$] is a function of the
concentration parameter ($ c \equiv R_\Delta/R_s$).  $R_\Delta$ is the
radius at which the total mass density is $\Delta$ times $\rho_c$.
For each model, the value of $\Delta$ is determined solely by the
redshift of the lens galaxy using the prescription of \citet{bry98},
which for our assumed cosmology allows $R_\Delta$ to be considered the
virial radius of the system.  For galaxies at $z \sim 0.6$, $\Delta$
is $\sim$ 140, and all galaxies in our lens sample have overdensity
values between $\Delta \sim 125$ and $\Delta \sim 160$.

By defining the virial radius, we are also able to further constrain
the concentration parameter by assuming a functional form given by:
\begin{equation}
c = \frac{9}{1+z}\left(\frac{M_{vir}}{8.12 \times 10^{12} h^{-1} M_\odot}\right)^{-0.14}
\label{eqn:numc}
\end{equation}
as found in numerical dark matter simulations
\citep{bul01,eke01,hoe05}, and where $M_{vir}$ is the total virial
mass.  While we assume no intrinsic scatter in the mass-concentration
relation for our initial analysis, we do consider the effects of
scatter, as well as other systematic effects that can implicitly alter
this relationship, in \S5.3.  Rather than attempt to constrain the
virial mass for each galaxy in our sample individually, we instead
choose to parameterize these masses as a function of V-band luminosity
($L_V$).  Specifically, we choose the form $M_{vir} = \tau_VL_V$,
where $\tau_V$ is defined to be the virial mass to light ratio.  This
then allows us to define the dark matter mass profile in terms of:
\begin{equation}
\Sigma_{\mathrm{NFW}}(R) = f(R,\Delta,c) \equiv f(R,z,M_{vir}) \equiv
f(R,z,L_V,\tau_V)
\label{eqn:dkfunc}
\end{equation}
As in the case of the stellar mass profile, we take $L_V$ from the
GALFIT models and, since the redshift is already known for each lens,
we once again are left with a model with a single free parameter, this
time $\tau_V$.

\subsection{Model Fitting}
\subsubsection{Mass Overdensity Models}
We fit the models to the observed $\Delta\Sigma$ profile, determining
the best-fit $M/L$ ratios associated with the lens galaxies through a
$\chi^2$ minimization procedure.  We compare the observed weak lensing
profile to the two-component models and average over contributions
from individual models in order to mimic the procedure used in the
weak lensing analysis.  The merit function for our $\chi^2$ procedure
is:
\begin{align}
&\nonumber\chi^2_{\mathrm{wl}} = \sum_{i=1}^{\mathrm{n_{bins}}}
  \left(\frac{1}{\sigma^2_{\Delta\Sigma,i}}\right)\bigg\{\Delta\Sigma(R_{i})~
  -
  \\&\frac{1}{\mathrm{N_{lens}}}\sum_{j=1}^{\mathrm{N_{lens}}}\left[\Delta\Sigma_{*,j}(R_{i},\Upsilon_V)
    + \Delta\Sigma_{\mathrm{NFW},j}(R_{i},\tau_V)\right]\bigg\}^2
\label{eqn:wlchi}
\end{align}
where $\Delta\Sigma(R_i)$ is the observed mass overdensity at radius
$R_i$ (Table \ref{tbl:mass}), $\sigma_{\Delta\Sigma,i}$ is the error
associated with that measurement, and $\Delta\Sigma_{*,j}$ and
$\Delta\Sigma_{{\rm NFW},j}$ are the stellar and dark matter mass
overdensity models evaluated at $R_i$ (using the parameters of lens
$j$), constructed according to Equation (\ref{eqn:delsig}).  This
optimization to the weak lensing data set is performed for the full
sample of lenses.

To take full advantage of our data set, we also incorporate the strong
lensing information.  We note that, for all strong lenses, the average
value of the mass surface density within the Einstein radius is
equivalent to the critical lensing density, $\Sigma_{\rm crit}$.  This
can be proven by noting that for a circularly symmetric lens, the
tangential reduced shear profile ($g_t$) can be written in the form:
\begin{equation}
g_t(R) = \frac{\overline{\Sigma}(<R) - \Sigma(R)}{\Sigma_{\mathrm{crit}} - \Sigma(R)}.
\label{eqn:gt}
\end{equation}
Because $g_t$ is, by definition, equal to 1 at $R_{\mathrm{Ein}}$ it
is a simple matter to show that $\overline{\Sigma}(<R_{\mathrm{Ein}})
= \Sigma_{\mathrm{crit}}$.

Thus, for all lenses for which we know both the lens and source galaxy
redshift, we can accurately determine the value of the critical
density of the system and further constrain the models.  The strong
lensing merit function is given by:
\begin{equation}
\chi^2_{\mathrm{sl}} = \sum_{i = 1}^{\mathrm{N_{lens}},z_s}\left[\frac{
\Sigma_{\mathrm{crit},i} - \overline{\Sigma}_*(R_{\mathrm{Ein},i}) - 
\overline{\Sigma}_{\mathrm{NFW}}(R_{\mathrm{Ein},i})}{\sigma_{\Sigma_{\mathrm{crit}},i}}\right]^2
\label{eqn:slchi}
\end{equation}
where $\Sigma_{\mathrm{crit},i}$ is the actual critical density of
lens i, $\sigma_{\Sigma_{\mathrm{crit}},i}$ is the uncertainty on that
value, $\overline{\Sigma}_*(R_{\mathrm{Ein},i})$ and
$\overline{\Sigma}_{\mathrm{NFW}}(R_{\mathrm{Ein},i})$ are the stellar
and dark matter average mass density models of lens $i$, evaluated at
the Einstein radius of that specific lens, and N$_{\rm{lens},z_s}$ is
the number of lenses for which the source redshift has been measured.

\subsubsection{Reduced Shear Models}
In addition to the mass overdensity model fit described in \S3.3.1, we
also fit models directly to our reduced shear data.  Although shear is
correlated with
%not entirely independent from 
mass overdensity, we do expect there to be a slight discrepancy
between the two methods: by creating models to fit reduced shear
directly, we no longer need to rely on the basic assumption of the
weak lensing regime that convergence is always small (i.e., $\kappa
\ll 1$).  Indeed, this assumption will break down as we approach the
Einstein radius, and to fully incorporate these smaller radii into the
model, we should account for the non-linear response of lensing in
this regime.  To that end, we create two new merit functions that
parallel the $\chi^2_{\mathrm{wl}}$ and $\chi^2_{\mathrm{sl}}$ used in
the mass overdensity fit.

For weak lensing data, we use the function:
\begin{align}
\nonumber\chi^2_{\mathrm{wl,shear}} =
&\sum_{i=1}^{\mathrm{n_{bins}}}\left(\frac{1}{\sigma^2_{g_t,i}}\right)\bigg\{
g_t(R_{i}) ~-
\\&\frac{1}{\mathrm{N_{lens}}}\sum_{j=1}^{\mathrm{N_{lens}}}g_{t_{\mathrm{mod}},j}
(R_{i},\Upsilon_V,\tau_V)\bigg\}^2
\label{eqn:wlchi_sh}
\end{align}
where $g_t(R_{i})$ is the observed reduced shear signal in a
given radial bin (Table \ref{tbl:mass}),
$\sigma_{g_t,i}$ is the measured uncertainty, and
$g_{t_{\mathrm{mod}},j}$ is the model combined (de
Vaucouleurs + NFW) reduced shear profile for lens $j$, given by:
\begin{equation}
g_{t_{\mathrm{mod}}} = \frac{(\overline{\Sigma}_* + \overline{\Sigma}_{\mathrm{NFW}}) - (\Sigma_* + \Sigma_{\mathrm{NFW}})}{\Sigma_{\mathrm{crit}} - (\Sigma_* + \Sigma_{\mathrm{NFW}})}
\label{eqn:shearmod}
\end{equation}
since individual components of reduced shear do not add linearly.
Once again, $\chi^2_{\rm{wl,shear}}$ is calculated for the full
sample.

For lenses where both $z_l$ and $z_s$ are known, we create
a strong lensing merit function, of the form:
\begin{equation}
\chi^2_{\mathrm{sl,shear}} = \sum_{i = 1}^{\mathrm{N}_{\mathrm{lens},z_s}}\left[\frac{
R_{\mathrm{Ein},i} - R_{\mathrm{Ein}}(g_{t_{\mathrm{mod}},i})}{\sigma_{R_{\mathrm{Ein}},i}}\right]^2
\label{eqn:slchi}
\end{equation}
where $R_{\mathrm{Ein},i}$ is the observed Einstein radius of lens
$i$, $\sigma_{R_{\mathrm{Ein}},i}$ is its measured uncertainty, and
$R_{\mathrm{Ein}}(g_{t_{\mathrm{mod}},i})$ is the model Einstein
radius {--} determined by inverting Equation (\ref{eqn:shearmod}) to
find the radius where the model reduced shear profile is equal to 1.

\section{Results}
\subsection{Total Mass Profile}

%-----------------------------------------------------------------------
\begin{figure*}
\plottwo{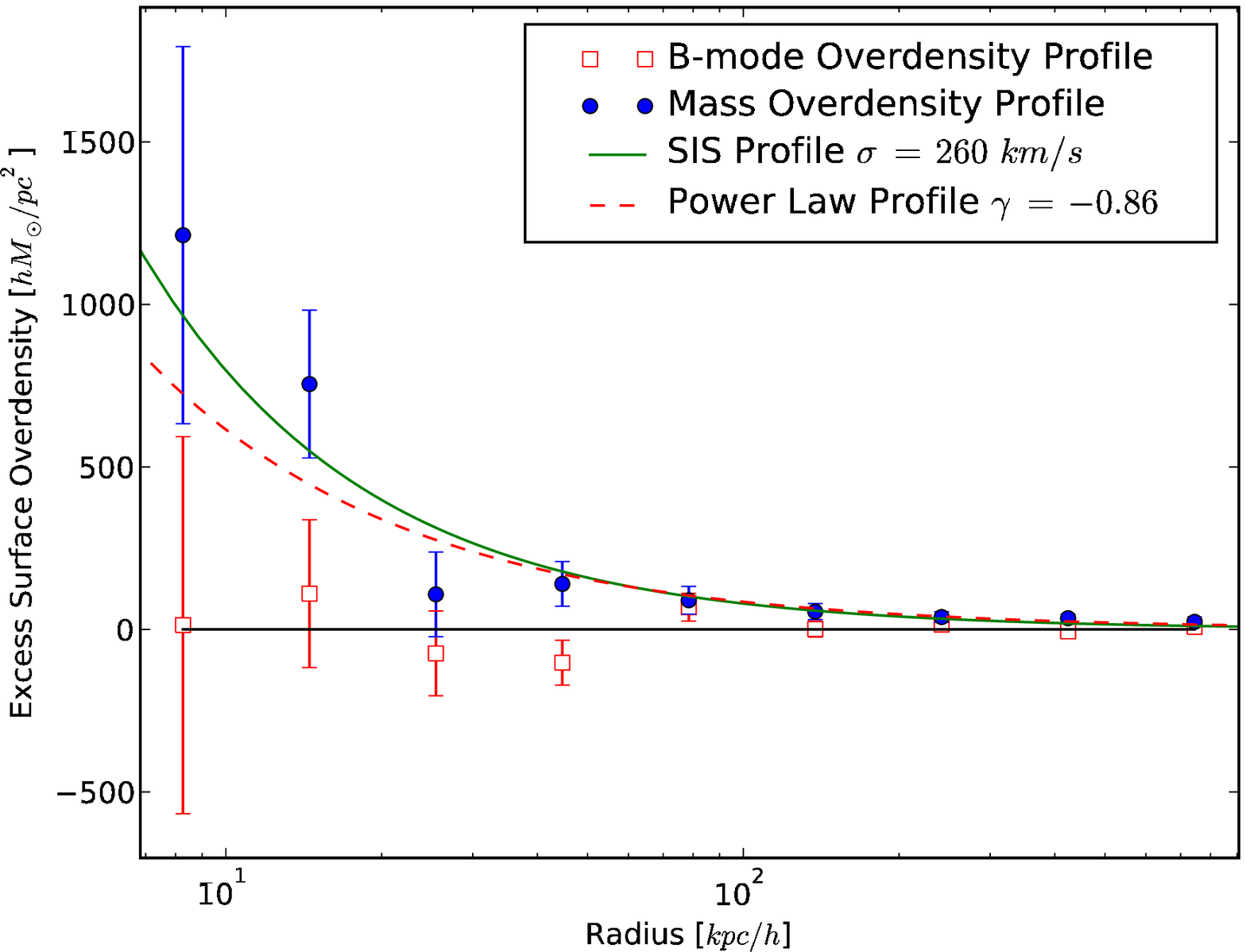}{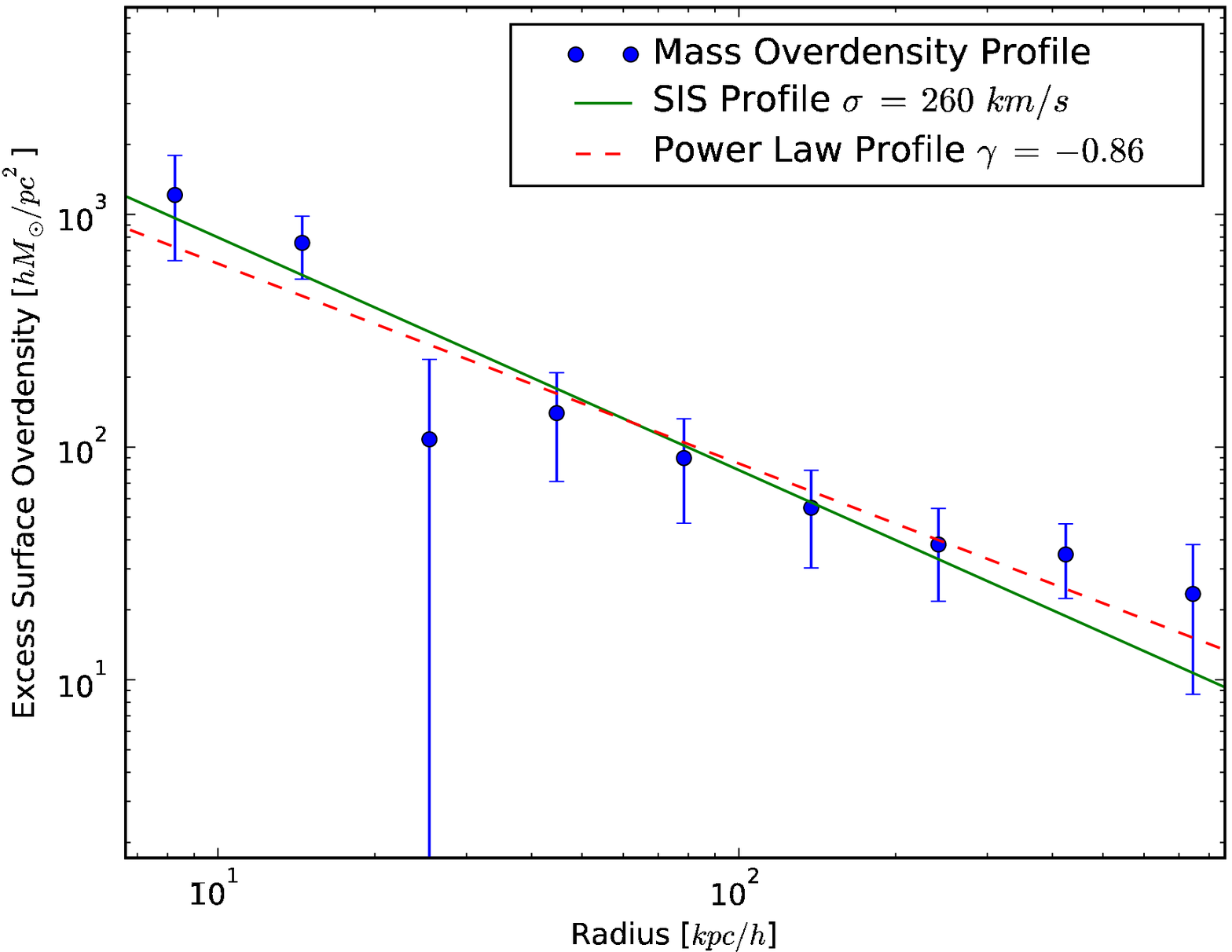}
\caption{Mass overdensity profile.  The best-fit generic power law
  model (red dashed line) is consistent with the data, and does not
  significantly deviate from the best-fit isothermal profile (green
  line).}
  
\label{fig:fmass}
\end{figure*}
%-----------------------------------------------------------------------

The observed mass overdensity profile can be seen in Figure
\ref{fig:fmass}. We can see a significant signal in the $\Delta\Sigma$
(E-mode) profile, while the $\Delta\Sigma_\times$ (B-mode) profile is
consistent with zero.  Measured values of both $\Delta\Sigma$ and
$\Delta\Sigma_\times$ are reported in Table \ref{tbl:mass}, along with
their associated shear values.  To measure the slope of the density
profile, we fit both a generic power law model ($\Delta\Sigma$
$\propto$ R$^{\gamma}$) and an isothermal model ($\Delta\Sigma$
$\propto$ R$^{-1}$) to the data (Figure \ref{fig:fmass}).  We find
that the best-fit power-law model ($\gamma = -0.86 \pm 0.16$) does not
deviate from the best-fit isothermal model, parameterized by a velocity
dispersion of $\sigma_v = (260 \pm 20)$ km s$^{-1}$, by more than
1-$\sigma$.  Given our uncertainties, both models are consistent with
the data, allowing us to conclude that the average mass overdensity
profile of our sample is approximately isothermal.
%-----------------------------------------------------------------------
\begin{deluxetable*}{rcrrrrrr}
\tablecaption{Measured Excess Surface Density and Tangential Shear. \label{tbl:mass}}
\tablehead{
\colhead{Galaxy Counts} & \colhead{Outer Bin Boundaries\tablenotemark{a}} & \colhead{$\Delta \Sigma$} & \colhead{$\Delta \Sigma_\times$} & \colhead{$\sigma_{\Delta \Sigma}$} & \colhead{$g_t$} & \colhead{$g_\times$} & \colhead{$\sigma_{g}$}\\ \colhead{} & \colhead{($h^{-1}$ kpc)} & \colhead{} & \colhead{} & \colhead{} & \colhead{} & \colhead{} & \colhead{}
}
\startdata
   4 	 &  11     &1213 &  13  &580  &0.231  &-0.010  &0.101 \\	
  23 	 &  19     & 755 & 110  &228  &0.124  & 0.023  &0.049 \\	
  73 	 &  33     & 108 & -73  &130  &0.024  &-0.010  &0.027 \\	
 233 	 &  57     & 140 &-102  & 69  &0.029  &-0.020  &0.015 \\	
 656 	 & 100     &  90 &  66  & 43  &0.017  & 0.014  &0.009 \\	
1851 	 & 176     &  56 &   2  & 25  &0.012  & 0.000  &0.005 \\	
4380 	 & 308     &  38 &  14  & 16  &0.007  & 0.004  &0.004 \\	
8287 	 & 541     &  33 &  -6  & 12  &0.008  &-0.001  &0.003 \\	
6997 	 & 949     &  23 &   8  & 15  &0.005  & 0.001  &0.003 \\	
\enddata

\tablenotetext{~}{A quantitative description of the mass overdensity
  profile (Figure \ref{fig:fmass}) for our lens sample.  Here, Galaxy
  Counts refers to the number of galaxies in each radial bin (given by
  projected radius).  The outer radius of each bin is shown in the
  radial bin boundaries column.  The next three columns show
  respectively the excess surface mass overdensity, ``B-Mode''
  overdensity, and uncertainty on each mass measurement in a given
  bin.  The final three columns show the associated shear quantities.}
\tablenotetext{a}{The inner radius for the innermost bin is 6 $h^{-1}$ kpc.}

\end{deluxetable*}
%-----------------------------------------------------------------------

\subsection{Luminous and Dark Matter Profiles}
\begin{figure}
\plotone{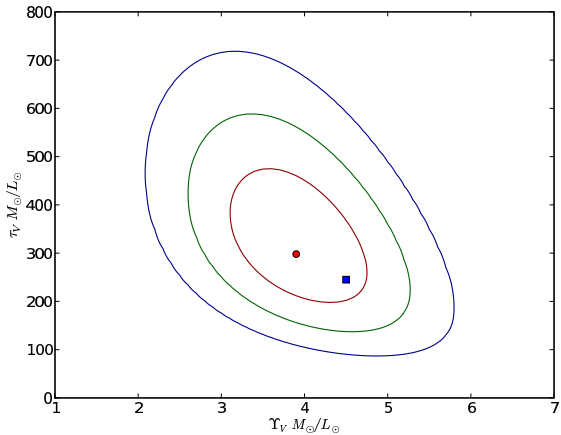}
\caption{Confidence contours for the best-fit $M/L$ ratios for mass
  overdensity models (red circle) and the associated 68\%, 95\%, and
  99.7\% confidence regions.  The best-fit value for the reduced shear
  analysis is shown by the blue square.  The results of the fits do
  not significantly disagree with one another.}
\label{fig:contours}
\end{figure}

We minimize the total merit function for each of the two approaches:
$\chi^2_{\mathrm{tot}} = \chi^2_{\mathrm{wl}} + \chi^2_{\mathrm{sl}}$
for the mass overdensity fit, and $\chi^2_{\mathrm{tot,shear}} =
\chi^2_{\mathrm{wl,shear}} + \chi^2_{\mathrm{sl,shear}}$ for the
reduced shear fit.  The best-fit $M/L$ parameters obtained with both
approaches can be seen in Figure \ref{fig:contours}.  The results from
the model fitting procedures disagree by less than one standard
deviation.  Thus, we feel we can use the results from the mass
overdensity model fit without worrying about inaccuracies in the
profile due to the simplifying assumptions made about shear.

%-----------------------------------------------------------------------
\begin{figure}
\plotone{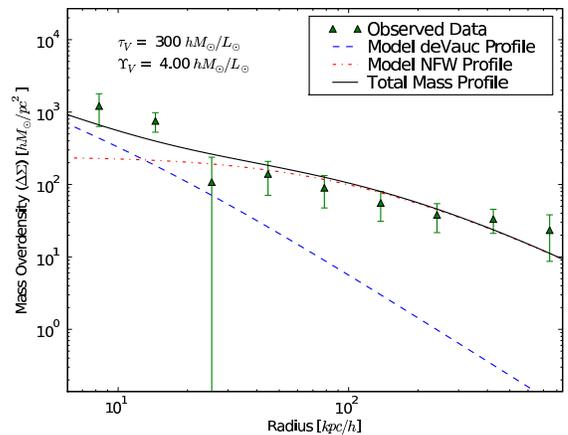}
\caption{Stellar and dark matter profiles generated by our
  best-fit mass-to-light ratios.  The blue dashed line represents the
  stellar de Vaucouleurs density profile, while the red dash-dotted
  line represents the dark matter NFW profile.}  
\label{fig:M/L_prof}
\end{figure}

%-----------------------------------------------------------------------
The mass overdensity profiles generated from our best-fit parameters
are shown in Figure \ref{fig:M/L_prof}, plotted along with the total
mass profile from our weak lensing analysis.  The figure shows the
contributions of the stellar (blue dashed line) and dark matter (red
dash-dotted line) mass profiles to the best-fit total mass profile
(solid black line).  The best fit to our data has $\chi^2_\nu = 0.98$,
indicating a good agreement between our model and the observed
overdensity profile.  As a comparison, the pure power-law fit in
\S{4.1} has $\chi^2_\nu = 0.81$, meaning that, given our errors,
neither the the 1- nor the 2-component model fit is strongly preferred
over the other.

We find a virial $M/L$ ratio $\tau_V = 300 \pm 90~h~M_\odot/L_\odot$ and
a stellar $M/L$ ratio $\Upsilon_V = 4.0 \pm 0.6~h~M_\odot/L_\odot$, for
an overall virial to stellar mass ratio of $M_{vir}/M_* = 75 \pm 25$.
Given the mean V-band luminosity of our sample, $\overline{L_V} = 4.5
\times 10^{10}~h^{-2}~ L_\odot$, this corresponds to average stellar
and virial masses of $\overline{M_*} = (1.8 \pm 0.3) \times 10^{11}
~h^{-1}~ M_\odot$ and $\overline{M_{vir}} = (1.4 \pm 0.4) \times
10^{13}~h^{-1}~ M_\odot$ respectively.  Furthermore, the average
virial mass we find corresponds to a typical virial radius of $332~
h^{-1}$ kpc and a typical NFW scale radius of $62~h^{-1}$ kpc.

Combining our mass ratio data with the assumed cosmology, we are also
able to estimate the stellar baryon fraction ($f_*$) present in our
lens sample, according to the relation:
\begin{equation}
f_* = \frac{M_{*}}{M_{vir}}\frac{\Omega_m}{\Omega_b}
\label{eqn:star_frac}
\end{equation}
where the quantity $\frac{\Omega_b}{\Omega_m}$ represents the total
baryon to dark matter ratio in the galaxy, and is given by the global
value of $\frac{\Omega_b}{\Omega_m} = 0.176 \pm 0.013$ \citep{spe07}.
For our sample, we find a stellar baryon fraction of $f_* = 0.075 \pm
0.030$, which is in excellent agreement with \citet{man06}, who find
$f_* = 0.055^{+0.015}_{-0.010}$ for their ``sm6'' bin, consisting of
early-type galaxies having $\overline{M_*} = 2.13 \times 10^{11}
~h^{-1} ~M_{\odot}$ and $\overline{M_{vir}} = 1.58 \times 10^{13}
~h^{-1} ~M_{\odot}$, and in good agreement with \citet{hey06} who find
$f_* = 0.10 \pm 0.03$.  The \citet{hey06} galaxy sample is somewhat
different from our own, in that it consists of both early- and
late-type galaxies.  The sample is dominated ($\sim75\%$ ) by early
types, though, and the authors note that adopting a more restrictive
selection criterion of S\'{ersic} index $n > 2.5$ does not change
their results.  The average stellar and virial masses of the
\citet{hey06} sample ($\sim 7 \times 10^{10} ~h^{-1} ~M_{\odot}$ and
$\sim 3 \times 10^{12} ~h^{-1} ~M_{\odot}$, respectively) are an order
of magnitude smaller than our own sample, making it difficult to
accurately compare results.  However, as the \citet{hey06} sample has
similar masses to the \citet{man06} ``sm4'' early-type sample, scaling
this value by $f_{*, \rm sm6} / f_{*, \rm sm4} = 0.33$ gives a result
of $f_* = 0.03 \pm 0.02$, which is still within $1.25-\sigma$ of our
value.

Finally, our $f_*$ value is in marginal agreement with \citet{fuk98},
who estimate a universal stellar baryon fraction of $f_* ~\sim~ 0.12$
for spheroids, using data acquired from the local universe.

%-----------------------------------------------------------------------
%-----------------------------------------------------------------------

\section{Discussion}
\subsection{Mass Profiles}
Figure \ref{fig:fmass} shows that the best-fit power law profile to
the total mass overdensity data between physical scales of $\sim$10
and $\sim$1000 $h^{-1}$ kpc has $\gamma = -0.86 \pm 0.16$, where
$\Delta\Sigma(R) \propto R^{\gamma}$.  This is consistent with an
isothermal ($\gamma = 1$) profile and, thus, is similar to the results
of G07, who found that the low-redshift SLACS lens sample also
displays a characteristically isothermal mass profile between $\sim$1
and $\sim$300 $h^{-1}$ kpc.  However, a closer inspection of the right
panel of Figure \ref{fig:fmass} reveals that we may be seeing some
flattening in the profile at large radii.  These points are located at
distances outside of the average virial radius of our lens sample.
Background galaxies located at these radii could thus be affected by
the dark matter halos of other massive galaxies nearby the lens, which
would explain the larger-than-expected shear signal in the profile.
Alternatively, since many of the galaxies in these bins lie near the
edge of the ACS field, we could be seeing excess ``shear'' due to CCD
edge effects.  Rerunning the profile analysis after excluding the
outermost two data bins, we find a best-fit profile $\gamma = -1.06
\pm 0.09$ that is much more consistent with an isothermal
distribution.

Even with the possibility of flattening at large radii, though,
combining the results of G07 with our results suggests that the
average mass overdensity profile of strong gravitational lenses is
consistent with an isothermal model over three orders of magnitude in
physical radius.  This result is in good agreement with previous mass
profile studies of non-lensing ellipticals \citep{she04,man06} between
$\sim 50 - 300 ~h^{-1}$ kpc, which would suggest that there is little
difference between the mass profiles of field strong lenses and their
non-lensing counterparts at moderate radii.

%-----------------------------------------------------------------------
\begin{figure}
\plotone{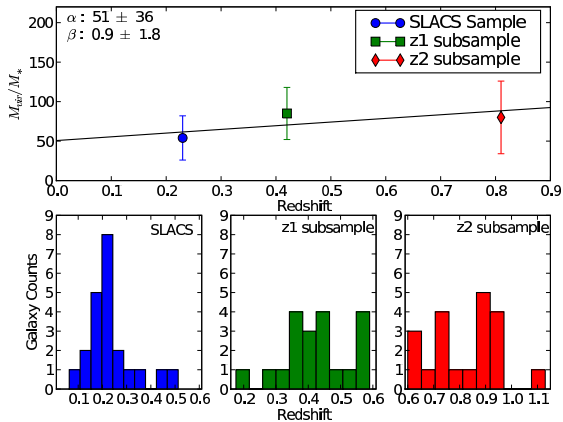}
\caption{Top: Evolution of the ratio of dark to luminous matter
  between SLACS and our lens sample, after splitting our sample into
  two separate redshift bins.  The best-fit slope of the data is shown
  as a solid line, with $\alpha = 51 \pm 36$ and $\beta = 0.9 \pm
  1.8$.  Bottom: Redshift histograms for each data point, showing the
  redshift distribution of lens galaxies within each bin.}
\label{fig:Mratio2}
\end{figure}
%-----------------------------------------------------------------------

%-----------------------------------------------------------------------
\begin{deluxetable}{lcccc}
\tablecaption{Mass Ratio Evolution. \label{tbl:evol}}
\tablehead{
\colhead{Sample} & \colhead{z} & \colhead{$L_V$} & \colhead{$M_{vir}/M_*$} & \colhead{$\sigma_{M_{vir}/M_*}$} \\
        \colhead{} & \colhead{} & \colhead{($10^{10}~h^{-2}~L_{\odot}$)} & \colhead{} & \colhead{}
}
\startdata
SLACS 	   &  0.23   &5.8   &54   &28 \\	
z1         &  0.42   &4.2   &85   &33 \\	
z2         &  0.83   &4.8   &80   &46 \\	
\enddata

\tablenotetext{~}{Data values associated with Figure \ref{fig:Mratio2}.
  The $z$ and $L_v$ values for each data point are, respectively, the
  average redshift and V-band luminosity of all lenses included in the
  bin.}
\end{deluxetable}
%-----------------------------------------------------------------------

Comparing the mean velocity dispersion of our lens sample
($\overline{\sigma_v}$ = 260 $\pm$ 20 km s$^{-1}$) to that of the G07
SLACS sample ($\overline{\sigma_v}$ $\simeq$ 248 km s$^{-1}$), and
noting the broad similarities between the slopes of their mass
profiles, it would appear that there is little difference between the
overall mass properties of strong lenses at $z = 0.6$ and the
properties of those at $z = 0.2$, although large uncertainties on both
profiles could easily hide any real evolutionary results.
Additionally, the best-fit $M/L$ ratios of our sample
($\tau_V = 300 \pm 90~h~M_\odot/L_\odot$, $\Upsilon_V = 4.0 \pm
0.6~h~M_\odot/L_\odot$) are not significantly different from those of
G07 ($\tau_V = 246 \pm 100~h~M_\odot/L_\odot$, $\Upsilon_V = 4.48 \pm
0.46~h~M_\odot/L_\odot$).

Considering the hierarchical merging model of galaxy formation, which
posits that the majority of the mass assembly of massive red
ellipticals (the dominant morphology of lensing galaxies) takes place
at redshifts $z \geq$ 2, it is not surprising that the differences in
total mass between these two samples is not dramatic.  More surprising
however is the fact that the individual $M/L$ ratios do not change, as
passive luminosity evolution could significantly affect the total
brightness of a galaxy over this redshift range.  We investigate the
effects of purely passive evolution on the stellar $M/L$ component in
two ways, back-evolving the G07 $\Upsilon_V$ result from $z = 0.2$ to
$z = 0.6$ using both the methods of \citet{tre01} (method 1) and
\citet{van01} (method 2).  After applying these relations, we find a
theoretical $M/L$ ratio of $\Upsilon_{V,z=0.6} = 3.1 \pm 0.3$ using
method 1 and $\Upsilon_{V,z=0.6} = 3.4 \pm 0.3$ using method 2.  Both
results agree with our measured $\Upsilon_V$ within the errors.  From
this, we conclude that the passive evolution and no-evolution models
are both plausible descriptions of stellar $M/L$ ratio evolution between
$\overline{z} \sim 0.2$ and $\overline{z} \sim 0.6$, given the current
size of our errors.  Future studies with larger sample sizes should be
able to place tighter constraints on these parameters, making it
easier to distinguish between the possible outcomes.

\subsection{Mass Ratio Evolution}
We also search for evidence of the evolution of mass ratio, looking
for any significant changes in the ratio of $M_{vir}/M_*$, which could
suggest evidence of a merger or satellite accretion event
\citep{con07}.  We do this by comparing our results to those of G07,
after splitting our sample into two redshift bins: subsample z1, where
$z < 0.6$ and subsample z2, where $z > 0.6$.  We fit a slope to
the data of the functional form:
\begin{equation}
\frac{M_{vir}}{M_*}(z) = \alpha(1+z)^{\beta}
\label{eqn:evol}
\end{equation}
The results of the fit can be seen in Figure \ref{fig:Mratio2}, and
the data are presented in Table \ref{tbl:evol}.  We find a best-fit
$\alpha = 51 \pm 36$ and a best-fit $\beta = 0.9 \pm 1.8$, which is
consistent with a no-evolution model ($\beta = 0$).  This result is in
agreement with the work of \citet{hey06}, who use weak-lensing alone
to probe the evolution of a sample of predominantly ($\sim$ 75\%)
early-type galaxies between $z=0.2$ and $z=0.8$.  After explicitly
assuming a linear growth factor, they find that $M_{vir}/M_*(z)$ =
$(34 \pm 11) + (31 \pm 35)z$.  In using weak-lensing alone however,
\citet{hey06} are unable to constrain stellar mass directly, since
weak-lensing is not able to probe mass at small radii where stellar
matter dominates.  Instead, they make an assumption about the nature
of stellar mass, using a Kroupa stellar IMF \citep{kro93} to scale the
average halo mass for their galaxy sample.  This is similar to
techniques used by other weak-lensing only mass ratio studies, such as
\citet{man06} who used a Kroupa IMF and \citet{hoe05} who used a
scaled Salpeter IMF \citep{sal55}.

Unlike these previous studies, our stellar mass values are obtained
purely from a luminosity-scaled fit to the mass data, without the need
to invoke a specific stellar IMF.  Indeed, this is the first such
study to attempt to quantify the evolution of the stellar mass ratio
without making any assumptions about the nature of the stellar IMF,
and the fact that we are able to obtain a consistent result, even with
such a small sample of lenses is promising.  It is likely that future
studies with larger lens samples will be able to place even better
constraints on stellar mass evolution, without having to worry about
possible systematic errors associated with the IMF, focusing instead
on only the lensing-inferred mass and total luminosity, which are both
directly determined from the data.

\subsection{Mass to Light Ratio Systematics}
In determining the best-fit stellar and virial $M/L$ ratios for our
lens sample, we determine statistical uncertainties from the
model-fitting procedure that are on the order of $\sim 20 - 30\%$.  In
this section, we investigate the impact of systematic uncertainties.
Specifically, we look at two such effects: varying the inner slope of
the dark matter profile, and modifying the dark matter
concentration-mass relation (CMR).

%-----------------------------------------------------------------------
\begin{figure}
\plotone{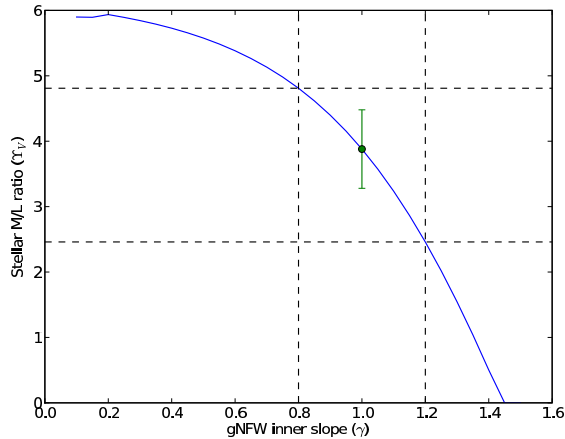}
\caption{Relationship between inner dark matter slope $\gamma$ and
  stellar $M/L$ ratio $\Upsilon_V$, showing a degeneracy between the
  parameters and demonstrating the fact that increasing the dark
  matter fraction in the core necessarily decreases the stellar mass
  fraction, assuming a constant total mass.  The data point represents
  our best fit $\Upsilon_V$ value assuming an NFW ($\gamma$ = 1)
  profile and the errors represent the 1-$\sigma$ confidence interval
  of that fit.  The dashed lines show the limits on $\Upsilon_V$ when
  $\gamma$ varies by 20\%.}
\label{fig:mlstar}
\end{figure}
%-----------------------------------------------------------------------

\subsubsection{Dark Matter Inner Slope}
When discussing the disentanglement of stellar and dark matter, we
have, to this point, specifically assumed a standard NFW profile for
the dark matter.  This has been done not only to take advantage of the
analytic form of the dark matter shear profile described in
\citet{wri00}, but also to compare our results to the numerous
previous studies that have assumed NFW dark matter profiles as well
(e.g. \citealt{hoe05,man06,hey06}; G07).  Recently however, studies
have shown that lensing data actually favor a dark matter profile with
a slightly modified inner slope ($\gamma$), often motivated by some
physical process such as adiabatic contraction
\citep[AC;][]{blu86,gne04} that affects the regions of a galaxy where
$R < R_{\rm s}$ \citep{jia07,gao08,sch09}.  Because this modification
only alters the distribution of the dark matter on small ($\lesssim 50
~h^{-1}$ kpc) scales for galaxy-scale masses, the total dark matter
mass and virial $M/L$ ratio derived from these new profiles remain
statistically consistent with the original NFW model.  However, this
is not the case for the derived stellar $M/L$ ratio ($\Upsilon$), as
even a change in $\gamma$ on the order of a few percent can
significantly impact $\Upsilon$, and hence the inferred stellar mass.

To gauge the systematic impact of altering $\gamma$ on our data, we
rerun the two-component stellar+dark matter analysis described in \S3,
replacing the standard NFW profile with a generalized NFW (gNFW)
profile \citep[e.g.][]{kee01}:
\begin{equation}
\rho(r) = \frac{\rho_s}{(r/r_s)^{\gamma}(1+r/r_s)^{3-\gamma}}
\label{eqn:gnfw}
\end{equation}
that is similarly projected into 2-dimensions (see \citealt{kee01} for
details).  While this prescription is somewhat different than applying
an AC model to a galaxy (AC also modifies the concentration parameter,
whereas altering the gNFW slope does not), the effect on the halo's
inner slope can be comparable: \citet{gne04} have demonstrated that,
assuming an initial NFW dark matter halo with a distribution of
baryons that condenses to form an elliptical galaxy, the final inner
slope of the dark matter halo ranges between $\gamma \sim 1.1$ and
$\gamma \sim 1.5$.

To be conservative, we vary $\gamma$ between 0.1 and 1.5, measuring
the best-fit V-band stellar $M/L$ ratio ($\Upsilon_V$) for each case.
The results can be seen in Figure \ref{fig:mlstar}.  We find that the
best-fit $\Upsilon_V$ is strongly dependent on $\gamma$, suggesting a
degeneracy between these two parameters that affects stellar mass in
the expected way: an increase in dark matter fraction in the core of
the galaxy (characterized by a steeper inner slope) results in a
smaller fraction of stellar mass (characterized by a lower stellar
$M/L$ ratio).

Assuming a fiducial uncertainty on $\gamma$ of 20\%, the variation in
$\gamma$ corresponds to a range in stellar $M/L$ ratio of $2.5 ~h
~(M_{\odot}/L_{\odot}) < \Upsilon_V < 4.8 ~h ~(M_{\odot}/L_{\odot})$.
Adding this systematic uncertainty to our previous estimate of
statistical error yields ($\Upsilon_V = 4.0 \pm 0.6 ~^{+0.8}_{-1.5} ~h
~M_{\odot}/L_{\odot}$).  As a comparison, \citet{jia07} measure a
$\sim$30\% decrease in stellar $M/L$ ratio by including the
\citet{blu86} AC model on a subsample of lower-redshift strong lenses,
in good agreement with the results we see when we increase the inner
slope of the NFW profile by $\sim$20\%.

\subsubsection{Concentration-Mass Relation}
By using a CMR to reduce the number of free parameters in our model
fitting routine, we are explicitly coupling the best-fit mass to the
observed shape of the mass density profile.  If instead we were to
make different assumptions about the CMR, this would lead to different
best-fit mass values, which ultimately would yield systematically
different stellar and virial $M/L$ ratios.  While we, like G07, have
assumed the \citet{bul01} CMR (Equation (\ref{eqn:numc})) in our
analysis, recent work involving higher resolution numerical
simulations \citep{mac07,mac08} have argued for an alternative CMR
that is less sensitive to halo mass and evolves more slowly with
redshift.  In particular, using the cosmological parameters from the
WMAP 5-year data set, \citet{mac08} determine a CMR of the form:
\begin{equation}
c = \frac{9.354}{(H(z)/H_0)^{2/3}}\left(\frac{M_{vir}}{1 \times 10^{12} h^{-1} M_\odot}\right)^{-0.094}
\label{eqn:maccio}
\end{equation}
where $H(z)$ is the Hubble parameter.  Comparing these CMRs for
galaxies with total mass of the order of our sample ($\sim 10^{13}
~h^{-1} ~M_{\odot}$), we find that the \citet{mac08} CMR yields a
concentration parameter that is $\sim15\%$ smaller than the
\citet{bul01} CMR at redshift $z = 0$, but (because of the slower
evolution with redshift) actually becomes $\sim10\%$ larger than the
\citet{bul01} CMR at $z = 0.6$, the median redshift of our sample.
Re-running our initial de Vaucouleurs + NFW model fit with this new
CMR, we find that the best fit $\tau_V$ and $\Upsilon_V$ parameters
vary by approximately 10\% and 5\% respectively, which is much less
than the statistical uncertainties presented in \S4.

In addition to completely modifying the CMR, we also investigate
systematics associated with the intrinsic scatter of the relation
itself.  Including the 1-$\sigma$ errors, \citet{bul01} measure a
$\sim40\%$ variation in halo concentrations for their sample of
$\sim10^{13} ~h^{-1} ~M_{\odot}$ halos, which we incorporate into our
analysis.  Specifically, assuming an extreme case where we increase
the concentration parameter of each galaxy in our sample by $\sim40\%$
(and the alternative case where we decrease each halo by $\sim40\%$),
we measure variations in $\tau_V$ and $\Upsilon_V$ of 25\% and 15\%
respectively.  Adding these uncertainties in quadrature with the
uncertainties associated with our choice of CMR, we therefore include
an additional $\sim15\%$ and $\sim25\%$ uncertainty on our best-fit
stellar and virial $M/L$ ratios due to the systematics of the CMR.

\subsubsection{Propagation of Systematic Uncertainties}
Ultimately, by investigating modifications to both the inner slope and
concentration parameter of the dark matter halo, we believe that we
are able to characterize the potential impact of important systematics
on our results, enabling us to propagate these uncertainties to all
other measured quantities that depend on them.  In particular, we find
a total systematic uncertainty of $\sim30\%$ on our best-fit
$\Upsilon_V$ (dominated by the systematics of the dark matter inner
slope) and a $\sim25\%$ uncertainty on our best-fit $\tau_V$, which
subsequently correspond to $\sim30\%$ and $\sim25\%$ uncertainties on
our derived stellar and virial masses.  Propagating these
uncertainties to the other derived parameters, we measure a $\sim40\%$
systematic uncertainty on both the mass ratio ($M_{vir}/M_*$), and
stellar baryon fraction $f_*$.

%-----------------------------------------------------------------------
\begin{figure}
\plotone{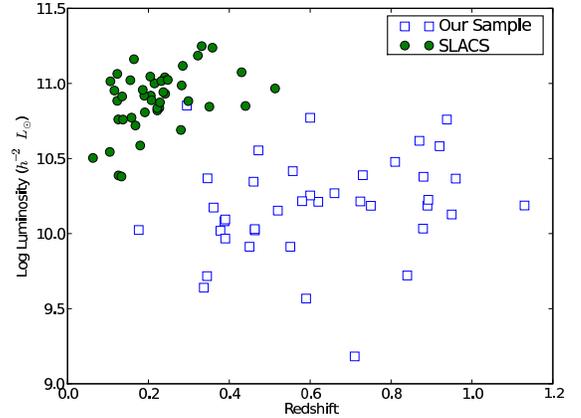}
\caption{Comparison of total luminosity (extinction corrected and
  passively evolved to $z = 0$) vs. observed redshift between the
  deep-exposure SLACS lens sample and the lenses used in this work.
  The SLACS lenses are more luminous than our sample.}
\label{fig:mlum}
\end{figure}
%-----------------------------------------------------------------------

\section{Conclusions and Future Work}
Taking advantage of the large density of background galaxies obtained
by relatively short ($\sim 1800$ sec) ACS exposures, we present the
results of a joint weak and strong lensing analysis of a sample of 41
massive elliptical galaxies at moderate ($z \sim 0.4 - 0.9$) redshift,
determining the mass overdensity profile over nearly 3 orders of
magnitude in physical radius.  Using a 2-component de Vaucouleurs +
NFW profile model, we are able to determine stellar and virial $M/L$
ratios for the sample, and disentangle the relative contributions of
stellar and dark matter from the total mass budget.  Furthermore, we
compare all of our results to those obtained from a subset of the
lower redshift SLACS sample ($\overline{z} \sim 0.2$), placing
constraints on the evolution of these mass properties over cosmic
time.  Our results can be summarized as follows:

\begin{itemize}

\item We present new redshift information for 7 COSMOS lenses: we find
  lens redshifts for J095930.93+023427.7, 0038+4133,
  J100140.12+020040.9, 0211+1139, and 0216+2955, a source redshift for
  0056+1226, and both lens and source redshifts for 0254+1430.

\item The mass overdensity profile of our sample has a best-fit
  power-law profile of $\gamma = -0.86 \pm 0.16$ and is consistent
  with an isothermal model between $\sim$ 10 $h^{-1}$ kpc and $\sim$
  1000 $h^{-1}$ kpc.

\item The best-fit stellar and virial $M/L$ ratios for our lens sample
  are given by $\Upsilon_V = 4.0 \pm 0.6 ~^{+1.1}_{-1.8}
  ~h~M_\odot/L_\odot$ and $\tau_V = 300 \pm 90 \pm 75
  ~h~M_\odot/L_\odot$ respectively.

\item Given our average sample luminosity of $\overline{L_V} = 4.5
  \times 10^{10}~h^{-2}~ L_\odot$, we find an average virial mass of
  $\overline{M_{vir}} = (1.4 \pm 0.4 \pm 0.3) \times 10^{13}~h^{-1}~
  M_\odot$ and an average stellar mass of $\overline{M_*} = (1.8 \pm
  0.3 ~^{+0.5}_{-0.8}) \times 10^{11}~h^{-1}~ M_\odot$.

\item We find an average mass ratio of $M_{vir}/M_* = 75 \pm 25
  ~^{+36}_{-27}$, which is consistent with the lower-redshift SLACS
  sample ratio of $M_{vir}/M_* = 54 \pm 28$.

\item Using our assumed cosmology, we convert mass fraction into a
  stellar baryon fraction, finding that $f_* = 0.075 \pm 0.030
  ~^{+0.030}_{-0.040}$.  This result indicates a low stellar formation
  efficiency in early-type galaxies, in agreement with previous
  results \citep{man06,hey06}.

\item By comparing our mass fraction data to the lower-redshift SLACS
  sample, we are able to place constraints on the evolution of the
  virial to stellar mass fraction of massive early-type galaxies.  We
  find that the quantity evolves as $M_{vir}(z)$/$M_{*}(z)$ = $(51 \pm
  36)(1 + z)^{(0.9 \pm 1.8)}$ over the last $\sim$ 7 Gigayears.  This
  is the first such study to place constraints on the evolution of
  $M_{vir}$/$M_*$ without invoking a specific IMF.

\end{itemize}

Overall, these results show that our method is a promising new
technique: with only a small sample of lenses (41 systems), were are
able to significantly measure the average mass overdensity profile and
mass-to-light ratios of the sample, allowing us to place constraints
on the mass evolution of early-type galaxies from $z \sim 0.8$ to $z
\sim 0.2$.  As always though, this work can be strengthened by
increasing the sample size.  While galaxy-scale strong lenses are
relatively rare today ($\sim$ 100 systems), future large-scale
observational surveys such as LSST ($\sim$ 10000 new lenses) and JDEM
\citep[$\sim$ 10000 new lenses;][]{mar05} are expected to dramatically
increase the number of known strong lenses.  Including these new
lenses in future work will greatly reduce the size of our present
statistical uncertainties.

In the near future however, we plan to increase our lens sample by
including other known deep space-based lens data from CASTLES that
have been imaged using the smaller HST Wide Field / Planetary Camera 2
(WFPC2).  Though the smaller area will limit our ability to probe out
to the large radii made available with the ACS sample, the longer
exposure times for the WFPC2 data yield background densities that are
comparable to those in the ACS sample, allowing us to place better
constraints on the inner mass overdensity profile by incorporating
these galaxies into our current lens sample.  We will also take
advantage of the HAGGLeS strong-lens search of bright red galaxies
\citep{mar09} which is expected to explore the whole of the HST ACS
archive, discovering $\sim$ 10 lenses per square degree of coverage.
We plan to further augment this study by combining our ACS sample with
the full ($\sim$ 40) long-exposure SLACS sample, which will allow us
to probe the differences in the mass properties of strong lenses
across a wider variety of samples and categories, such as morphology,
luminosity, and finer redshift slices.  We are particularly interested
in segregating the lens sample by total luminosity, since the SLACS
lenses appear to be more luminous than their moderate-redshift
counterparts when passively evolved to $z = 0$ (Figure
\ref{fig:mlum}), suggesting that the SLACS sample may have evolved
from a slightly different population of galaxies.

Of course, by increasing the sample size of lenses and thus reducing
the statistical errors associated with these measurements, correcting
systematic uncertainties will become much more important.  In future
studies we hope to improve our control of systematics in several ways.
Foremost is the ability to measure accurate redshifts for both the
lens and source galaxies of strong lens systems, as these measurements
are crucial for accurate lens modeling.  While some future survey
instruments should provide redshift information (either
spectroscopically or photometrically) as part of their normal
operation, it is still important to improve the redshift information
for currently known lenses, and we plan to obtain more spectroscopic
and photometric redshift data in the near future.  Additional
systematics controls could come in the form of improved profile
modeling, relaxing the strict de Vaucoulers + NFW mass profiles in
favor of a more general free-index S\'{e}rsic profile for the stellar
mass and a freely-varying gNFW profile for the dark matter.
Similarly, improved strong lens mass modeling should be utilized in
the future, increasing our ability to distinguish between various mass
profiles at very small scales, as well as improving the precision of
our stellar mass fraction measurements.  Finally, we could improve
galaxy selection and weak lensing measurements by using HST data
obtained through other filters, allowing us to determine photometric
redshifts for many of the background source galaxies.  We note however
that this should not be a dominant source of systematic error, as we
have shown previously that altering the background galaxy redshift
distribution does not significantly change the results of any of our
fitted parameters.

\begin{acknowledgements}
DJL thanks Brian Lemaux, Jim Bosch, Ami Choi, and Will Dawson for
useful discussions regarding this project.  DJL thanks Aaron Dutton
for providing insight into the discussion of dark matter systematics
and concentration-mass relations.
DJL and CDF acknowledge support from program \#HST-AR-11246, provided
by NASA through a grant from the Space Telescope Science Institute,
which is operated by the Association of Universities for Research in
Astronomy, Incorporated, under NASA contract NAS5-26555.
DJL, CDF, and MWA acknowledge support from NSF-AST-0909119.
PJM was given support by the TABASGO
foundation in the form of a research fellowship. The HAGGLeS legacy
archive project was funded by HST grant 10676.
MB acknowledges support from NASA through Hubble Fellowship grant
\#~HST-HF-01206 awarded by the Space Telescope Science Institute.
TS acknowledges support from NWO.
Some of the data presented herein were obtained at the W. M. Keck
Observatory, which is operated as a scientific partnership among the
California Institute of Technology, the University of California, and
the National Aeronautics and Space Administration. The Observatory was
made possible by the generous financial support of the W. M. Keck
Foundation.  The authors wish to recognize and acknowledge the very
significant cultural role and reverence that the summit of Mauna Kea
has always had within the indigenous Hawaiian community.  We are most
fortunate to have the opportunity to conduct observations from this
mountain.
\end{acknowledgements}

\begin{appendix}
\section{Weak Lensing Analysis}
\subsection{Object Detection and Shape Measurement}
This study utilizes the
IMCAT\footnote{http://www.ifa.hawaii.edu/$\sim$kaiser/imcat/} software
suite, which is based on the well-known algorithm of Kaiser, Squires,
\& Broadhurst (1995, hereafter KSB), as well as a series of Perl and
Python scripts which apply the PSF correction scheme of \citet{sch07}
and mass profile analysis of G07.  Using both the catalogs and their
associated science images, our pipeline first invokes the IMCAT object
detection program \textit{hfindpeaks} on the images and
cross-correlates the results with the related SExtractor catalogs, in
order to reduce the number of false detections.  Objects detected in
both SExtractor and \textit{hfindpeaks} are convolved with a Gaussian
smoothing kernel in order to regularize their shapes and sizes and
also to reduce shape noise; the width of an object's smoothing kernel
($r_g$) is given by the object's half light radius as determined by
SExtractor.  Once an object has been smoothed, the local sky
background is calculated and subtracted from its overall intensity.
After background subtraction, the object's magnitude is determined
using a magnitude zero point taken from the ACS
website\footnote{http://www.stsci.edu/hst/acs/}.

Finally, the script determines a centroid for each object, and
calculates the quadrupole moments of intensity, given by
\begin{equation}
Q_{ij} = \int_{r < 4r_g} d^2\theta W(\vec{\theta}) \theta_i\theta_j I(\vec{\theta}),~~i,j\in{x,y}
\label{eqn:quad}
\end{equation}
where the integral is taken from the centroid location to
$4 \times r_g$.  Here, $I(\vec{\theta})$
represents the intensity of an object in a given pixel, and
$W(\vec{\theta})$ is a weight function, which for this study is a
circular Gaussian with radius $r_g$, matching the filter used for
object detection.  These moments of intensity are then used to
calculate ``ellipticity polarizations'':
\begin{align}
e_1 = \frac{Q_{xx} - Q_{yy}}{Q_{xx} + Q_{yy}} \label{eqn:e1}\\
e_2 = \frac{2Q_{xy}}{Q_{xx} + Q_{yy}}
\label{eqn:e2}
\end{align}
which represent the components of the observed shape of the object.
In addition, the ``smear'' and ``shear'' polarizability pseudo-tensors
\citep[$P^{sm}$ and $P^{sh}$ respectively; see e.g. KSB, or the
  updated definitions in][]{hoe98} are also calculated, which are used
to correct shape measurement errors due to PSF distortion.% (\S 3.3).

\subsection{Charge Transfer Inefficiency Correction}

%-----------------------------------------------------------------------
\begin{figure}
\plotone{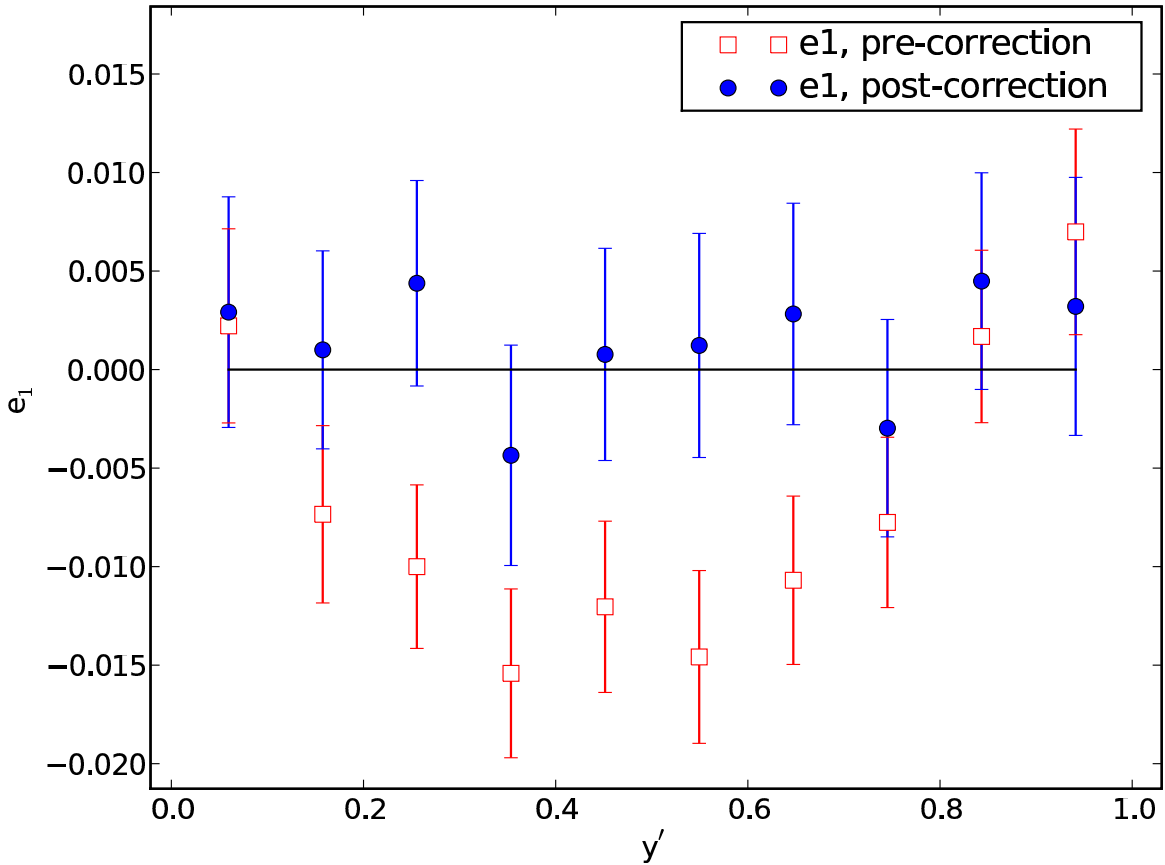}
\caption{Average $e_1$ component of ellipticity as a function of
  relative y pixel ($y^{\prime}$), with $y^{\prime} = 0$ [$y^{\prime}
    = 1$] representing the bottom [top] of the ACS field of view.  The
  effect of CTE degradation is seen as an average nonzero ellipticity
  in the $-e_1$ direction, and is seen most prominently in our
  uncorrected ellipticity bins (red squares) near the region farthest
  away from the CCD readout registers ($y^{\prime} = 0.5$), as we
  expect.  Our correction for CTE degradation (blue circles)
  removes nearly all of the spurious stretching.}
\label{fig:cte}
\end{figure}
%-----------------------------------------------------------------------

Before we analyze the weak lensing signal, we must correct for any
systematic distortions of the galaxy shapes.  The first such
systematic that we address is distortion due to the degradation of
charge transfer efficiency (CTE) of the ACS CCDs.  Defects that arise
during the lifetime of the CCD can cause some charge to be delayed and
transferred to other pixels, resulting in a trail of charge following
an object.  In the ACS, the readout direction is vertical.  Thus, this
trail will induce a spurious stretching in the $-e_1$ direction.
Since CTE degradation alters the shapes of galaxies in the same
direction, the effect can mimic the gravitational shear signal caused
by weak lensing.  The spurious polarization induced by CTE degradation
is typically on the order of 1\%, and therefore it most strongly
affects the outer regions of a galaxy-galaxy shear profile (where the
gravitational shear drops below 1\%).  If left uncorrected, this
spurious signal can alter the inferred total mass of the galaxy.

The ``strength'' of an object's tail is affected by several factors,
including object brightness, size, and location along the CCD readout
path.  This makes for an inherently non-linear problem that cannot be
corrected by a simple model. Instead, we use the empirical
prescription of \citet{rho07} devised for the COSMOS program.
Following G07, we modify the original equation to be compatible to our
data, although we use a slightly different prefactor which accounts
for combining the IMCAT definition of SNR with the SExtractor measured
half-light radius (G07 obtain both parameters from IMCAT, whereas
\citet{rho07} use SExtractor exclusively):
\begin{align}
\nonumber e_{1,\mathrm{CTE}} &= -6.5 \times 10^{-4}\left(\frac{1}{2} - \left|\frac{1}{2} -
y'\right|\right)(\mathrm{SNR})^{-0.9} \\&\times (\mathrm{MJD} -
52333)\left(\frac{r_h}{0.18\arcsec}\right)^{-0.1}
\label{eqn:cte}
\end{align}
which modifies the $e_1$ component of each object in a catalog.  In
this equation, $r_h$ represents the half-light radius of the object
(as determined by IMCAT), and $y^\prime$ is the normalized
y-coordinate of the object (equal to 0 at the bottom of an ACS field
and 1 at the top), defined such that the correction for the CTE
degradation is maximized near the ACS chip gap ($y^\prime \sim 0.5$),
where the pixels are the farthest away from the readout registers.
The time-dependent factor takes into account the fact that the effects
of CTE degradation have steadily worsened over the lifetime of the
ACS.

Figure \ref{fig:cte} shows the effect of CTE correction on our data
sample, and from the figure we can see that the average corrected
galaxy shape ($e_{1,\mathrm{corr}} = e_1 - e_{1,\mathrm{CTE}}$) is no
longer spatially dependent, and is very nearly zero throughout the
entire field.  In addition, our CTE correction scheme is able to
correct specifically chosen subsets of the galaxy population as well:
after separating the source galaxies into three distinct magnitude
bins (m $<$ 25; 25 $<$ m $<$ 26; m $>$ 26), we find that the
CTE-corrected galaxies in each bin show a mean shape
$e_{1,\mathrm{corr}} = 0$, again with no spatial dependence across the
field.

To test the systematics associated with our CTE correction scheme, we
apply an additional 20\% CTE shape correction to every background
galaxy, regardless of size, brightness, or position on the detector.
We perform a full weak lensing analysis and 2-component model-fit
(\S3) on these data, and find that the best-fit stellar and virial
$M/L$ ratios systematically vary from the values inferred from the
original data (presented in \S4) by $\sim 2\%$ and $\sim 15\%$
respectively, which are much smaller than the presented statistical
errors.

\subsection{PSF Correction}
We next perform a PSF correction to remove any shape distortions of
the background galaxies due to the telescope, leaving only those
distortions caused by gravitational shear.  Our correction scheme
applies the method of \citet{sch07}, which utilizes the KSB formalism
along with modifications developed by \citet{hoe98} and \citet{erb01}.

While it should be possible to map the spatial variations of the PSF
by directly measuring the shapes of the stars found in the data, we
find that this is not the case for our lens sample, as there are too
few stars per exposure to accurately sample the whole field.  Instead,
to increase our star count we use a series of well-sampled, low
galactic latitude stellar field exposures to generate a series of
polynomial PSF models.  There is some concern in fitting a model PSF
to our data instead of deriving it empirically, as the ACS PSF is
temporally variable.  However, this variability is primarily a
function of a single parameter: changes in the separation of the
primary and secondary HST mirrors due to thermal breathing
\citep{kri03,rho07}.  By using a total of 181 exposures taken over the
course of several months, our stellar-field PSF models span a wide
range of mirror separations, allowing us to compile a nearly
continuous database of varying PSF patterns that can be used to
correct our science images.

To create each model, we measure the anisotropy components ($q^*_{\rm
  i}$) and polarization tensor ratio ($T^*$) {--} a parameter required
for PSF correction that is related to the PSF width {--} as a function
of position, using the stars in a given stellar field.  The measured
quantities are given by:
\begin{align}
q^*_{\rm i} &= \left(P^{sm*}\right)^{-1}_{\rm ij}e^*_{\rm  j} \label{eqn:psf_ani}\\ 
T^* &= \frac{\rm{Tr}\left[P^{sh*}\right]}{\rm{Tr}\left[P^{sm*}\right] \label{eqn:psf_size}}
\end{align} 
where $e^*_{\rm j}$ are the measured stellar ellipticity components,
and $P^{sh*}$ and $P^{sm*}$ are the stars' shear and smear
polarization tensors, respectively.

Model stars are initially selected by simple cuts in size-magnitude
space.  A 0.8 pixel wide cut in half-light radius, centered on the
stellar locus, is used to separate stars from galaxies.  A bright-end
magnitude cut is used to remove stars with saturated pixels and a SNR
cut is used to remove stars that are too noisy to be accurately
measured.  From this initial star selection we generate 3rd-order
polynomial models for the $q^*_{\rm i}$ components, and a 5th-order
polynomial model for $T^*$.  To improve the model, we evaluate each
polynomial component at the coordinates of the stars, and compare the
inferred values to the actual measured quantities.  If, for any given
model component, the observed value disagrees with the model by more
than 2-$\sigma$, the star is considered an outlier (because of shape
noise) and is removed from the sample.  Once all outliers are removed,
we regenerate the polynomial models with the remaining stars.  This
process is repeated iteratively until the number of stars remains
stable, at which point we generate a final set of polynomials that
represent the final PSF model for that stellar field.

For each science exposure, we compare the measured PSF of each star in
the field to the model PSFs, and determine the best-fit model by
calculating the reduced $\chi^2$ statistic given by:
\begin{equation}
\chi^2_j=\displaystyle\sum_{i =
  1}^{\rm{N_{stars}}}\frac{\displaystyle\left[\rm{PSF}^{*}_i -
    \rm{PSF^{mod}}_j(x_i,y_i)\right]^2}{\rm{N_{stars}}}
\label{eqn:chisq} 
\end{equation}
where $\rm{PSF}^{*}_i$ is the measured PSF of star $i$, and
$\rm{PSF^{mod}}_j(x_i,y_i)$ is the model PSF of stellar template $j$,
evaluated at the location $\rm (x_i,y_i)$ of the real star.

Once best-fit PSF models are determined for all exposures associated
with a composite science image, the coordinates of each model are
registered to the science image, and the models are combined by an
exposure-time weighted average to create a final ``master'' PSF model
used to correct each background galaxy, given by:
\begin{equation}
\rm{PSF^{combo}}(x,y) = \frac{\displaystyle\sum_i^{\mathrm{N_{exp}}}~t_i~\rm{PSF_{i}(x,y)~\Delta_i}}{\displaystyle\sum_i^{\mathrm{N_{exp}}}~t_i \Delta_i}
\label{eqn:psfcombo}
\end{equation}
where $\rm t_i$ is the exposure time of a given exposure in the
composite science image, and $\Delta_{\rm i}$ is equal to 1 if the
galaxy falls within the boundaries of the WFC chips in the exposure
(not always true, due to dithering), and 0 otherwise.

%-----------------------------------------------------------------------
\begin{figure}
\plotone{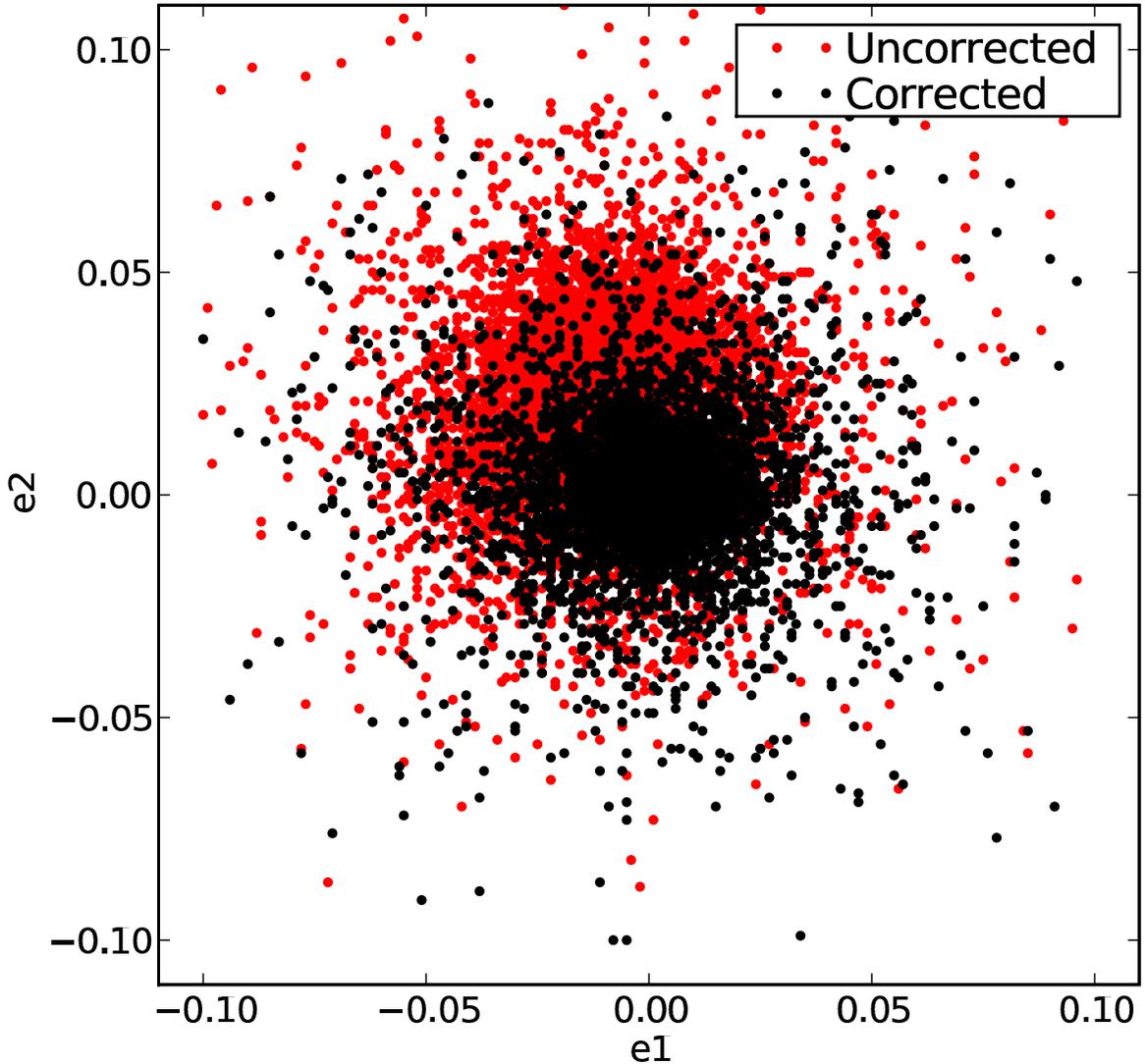}
\caption{Result of PSF correction on all stars of our science
  images.  Pre anisotropy-corrected shapes are plotted in red, while
  post-corrected shapes are shown in black.  We can clearly see that
  the stars have become more circular after applying the combined PSF
  models.}
\label{fig:psf}
\end{figure}
%-----------------------------------------------------------------------

As a test of our PSF correction, we plot the shapes of all stars found
in our science images, both pre- and post-correction, (Figure
\ref{fig:psf}).  We do see an improvement: the stellar ellipticity
components become more circular ($\overline{e_{1,pre}} = -0.0089$,
$\overline{e_{1,post}} = 0.0001$; $\overline{e_{2,pre}} = 0.0124$,
$\overline{e_{2,post}} = 0.0015$), and show a mild reduction in spread
($\sigma_{e_{1,pre}} = 0.022$, $\sigma_{e_{1,post}} = 0.019$;
$\sigma_{e_{2,pre}} = 0.021$,$\sigma_{e_{2,post}} = 0.016$).  By
taking into account effects due to the PSF (and effects due to CTE
degradation in the previous section) we are able to significantly
reduce the dominant sources of systematic error involved in the weak
lensing analysis.

\subsection{Data Cuts}
To improve the overall signal, we apply a series of data cuts to our
master catalog.  First, we exclude objects that have a half-light
radius smaller than 3 pixels, as many of these objects are ``bad''
data objects such as stars or cosmic rays, which are not affected by
gravitational shear and only serve to dilute the overall shear signal.
We also reject objects with a half-light radius larger than 14 pixels,
and ``bright'' objects ($m_{F814W} < 23$), as these objects are likely
to be foreground interlopers.  Additionally, we apply a conservative
cut to galaxy shapes, rejecting any object with $\left|g_i\right| > 2$
(where the $g_i$ are the CTE + PSF corrected ellipticity components)
as being unphysical or poorly shape-corrected in some way.  Finally,
we apply a general SNR cut to the data, rejecting any remaining object
with an IMCAT significance ($\nu$) parameter less than 16; (as a
comparison, this corresponds to an SNR $<$ 4 cut using the SNR
determination of \citet{erb01}).  After cutting the data, we are left
with a total background galaxy density of $\sim 65$ galaxies
arcmin$^{-2}$ (as compared to the $\sim 100$ galaxies arcmin$^{-2}$
measured prior to cutting) which is high enough to measure the weak
lensing signal around the galaxy stack.

\subsection{Catalog Stacking}
After correcting for PSF and CTE effects for objects in an individual
science field, we can apply the standard weak lensing formalism
\citep[see e.g.,][]{sei97,bar01,wit02} to the field and determine an
average mass profile from a stack of fields.

To properly combine each lens field into a cohesive stack, we first
apply a shift to the coordinates of each science field, such that the
origin in each field corresponds to the location of the strong lens
system.  Because the lenses in our sample cover a range of redshifts,
we convert each background galaxy's angular separation from the lens
to a physical one, so that we can more easily stack galaxy fields
together in catalog space.  Finally, once all objects have been
repositioned in terms of physical distances relative to a common
center, we combine everything into a master galaxy catalog that
contains information on galaxy position and corrected ellipticity, and
information on the strong lens field from which it was initially
drawn.  In addition, geometric information about the strong lenses
themselves are also included, which are critical in determining the
physical mass profile at small scales.

\subsection{Shear and Mass Profiles}
Since our master galaxy catalog contains information on both position
and corrected ellipticity, we can combine these parameters to develop
two radial shear profiles: a tangential ``E-mode'' profile given by:
\begin{equation}
\gamma_t(R) = -\left(\displaystyle\sum_{i=1}^{n_{gals,R}} \gamma_{1,i}~\cos(2\phi) + \gamma_{2,i}~\sin(2\phi)\right)
\label{eqn:tanshr}
\end{equation}
and a 45$\degr$ rotated ``B-mode'' profile, given by:
\begin{equation}
\gamma_\times(R) = -\left(\displaystyle\sum_{i=1}^{n_{gals,R}} -\gamma_{1,i}~\sin(2\phi) + \gamma_{2,i}~\cos(2\phi)\right)
\label{eqn:crsshr}
\end{equation}
where $n_{gals,R}$ refers to the number of galaxies located within a
radial bin centered at radius $R$ and $\gamma_{1,i}$ and
$\gamma_{2,i}$ represent the shear components of galaxy $i$,
equivalent to the components of reduced shear ($g_i$) in the limit of
weak lensing.  We calculate $\gamma_t(R)$ and $\gamma_\times(R)$ in
concentric circular annuli around the common center of our stacked
lens sample.  For a system with a single lens plane (or a stacked,
coincident lens plane) the lensing signal measured with respect to the
center of mass and averaged over all background galaxies should be
contained entirely within the E-mode profile while the B-mode should
vanish.  Thus, by computing these two shear profiles, we can see
information regarding the radial characteristics of the shear field
from $\gamma_t$, as well as a check on systematic errors from
$\gamma_\times$.

Next, we convert shear into a differential surface mass overdensity,
$\Delta\Sigma$, by scaling the tangential shear profile by the
critical lensing density $\Sigma_{\rm crit}$:
\begin{equation}
\Sigma_{\rm crit} = \frac{c^2}{4\pi G} \frac{D_s}{D_l D_{ls}}
\label{eqn:scrit}
\end{equation}
a measurement of surface mass density that encodes the angular
diameter distances between the observer and the lens ($D_l$), the
observer and the source ($D_s$), and the lens and the source
($D_{ls}$).  The $\Delta\Sigma(R)$ profile is thus given by:
\begin{equation}
\Delta\Sigma(R) \equiv \overline{\Sigma}(<R) - \Sigma(R) = \Sigma_{\rm crit}\gamma_t(R)
\label{eqn:delsig}
\end{equation}
where $\overline{\Sigma}(<R)$ represents the average surface mass
density enclosed within projected radius R.  Since $\Sigma_{\rm crit}$
is a function of both lens and source redshifts ($z_l$ and $z_s$,
respectively), in principle we must construct a unique $\Sigma_{\rm
  crit}(z_l,z_s)$ for every source galaxy that is used in the mass
overdensity profile.  In practice, however, this is impractical since
a large number of background galaxies do not have measured redshifts.
Instead, as in G07, we define an average critical density for each
lens field:
\begin{equation}
\Sigma^{\prime}_{\rm crit} = \left(\frac{c^2}{4\pi G}\right)\left[\frac{1}{D_l\overline{w}(z_l)}\right]
\label{eqn:avscrit}
\end{equation}
where $\overline{w}(z_l)$ is a weighted factor of ($D_s$/$D_{ls}$),
designed to fully account for background source redshift distribution.
It is given by:
\begin{equation}
\overline{w}(z_l) = \int_{z_l}^{\infty}dz_s\left(\frac{dn(z_s)}{dz_s}\right)\frac{D_{ls}(z_s)}{D_s(z_s)}
\end{equation}
where $dn(z_s)/dz_s$ is the redshift distribution of background source
galaxies.  Since our image exposure times and data cuts are similar to
G07 we use their redshift distribution, extrapolated from COSMOS:
\begin{equation}
\frac{dn(z_s)}{dz_s} = \frac{b}{z_0\Gamma(a/b)}e^{-(z_s/z_0)^b}(z_s/z_0)^{a-1}
\end{equation}
with $z_0$ = 0.345, $a$ = 3.89, and $b$ = 1.  From these parameters,
we calculate a mean background source redshift of $\overline{z_s} =
1.34$. A full list of $\overline{w}(z_l)$ parameters can be seen in
Table~\ref{tbl:lens}.

We next compute an estimator of $\Delta\Sigma(R)$:
\begin{equation}
\Delta\Sigma(R) = \frac{\displaystyle\sum_{l=1}^{N_{\rm lens}} \left[\Sigma'^{-1}_{{\rm crit},l}\displaystyle\sum_{i=1}^{N_{{\rm source},l}}\gamma_{t,i}~\sigma_{e,i}^{-2}\right]}{\displaystyle\sum_{l=1}^{N_{\rm lens}} \left[\Sigma'^{-2}_{{\rm crit},l}\displaystyle\sum_{i=1}^{N_{{\rm source},l}}\sigma_{e,i}^{-2}\right]}
\label{eqn:massest}
\end{equation}
where $N_{{\rm source},l}$ is the number of background galaxies
associated with lens field $l$, $\Sigma'_{{\rm crit},l}$ is the
average critical density for lens field $l$, and $\sigma_{e,i}$ is the
uncertainty assigned to the tangential shear estimate of galaxy $i$,
which we take to be the shape uncertainty as defined in
\citet{gavSou07} to be proportional to a galaxy's shape measurement
error added in quadrature with the intrinsic shape noise of all
galaxies ($\sigma_0 = 0.3$) to prevent over-weighting.  With this
conversion, we are able to transform shear to mass, and develop a mass
overdensity profile for our data.

As a check on the systematics of our assumed redshift distribution, we
vary the parameters of the distribution, calculating a unique set of
\{$a,b,z_0$\} parameters for each lens field.  We do this by measuring
the observed F814W magnitude distribution of the galaxies in the field
and convert this to a redshift distribution, using the method
described in \citet{sch07}.  After doing this, we recalculate the
$\Sigma^{\prime}_{\rm crit}$, $\overline{w}(z_l)$, and
$\Delta\Sigma(R)$ estimator values for the entire sample, and repeat
the weak lensing analysis with these updated values.  We find that,
regardless of the choice of redshift distribution parameters, the
best-fit model parameters determined here (as well as those described
in \S 3) vary by $\la 7 \%$.  Since this is smaller than the
statistical uncertainty of our measurements, we conclude that
uncertainties in the redshift distribution are negligible for our
purposes.
\end{appendix}

\end{document}